\documentclass[12pt,preprint,aps,showpacs,floatfix]{revtex4}
\usepackage{epsfig,colordvi}
\usepackage{color}
\usepackage{here}

 \def\as{$\alpha_S\ $}
 \def\T{\textstyle}
 \def\l{\left}
 \def\r{\right}
 \def\nf{n_{\!f}}

 \def\mds{\tilde{m}_D}

 \def\be{\begin{equation}}
 \def\ee{\end{equation}}
 \def\bea{\begin{eqnarray}}
 \def\eea{\end{eqnarray}}
 \def\bean{\begin{eqnarray*}}
 \def\eean{\end{eqnarray*}}
 \def\gsim{\mathrel{\rlap{\lower0.2em\hbox{$\sim$}}\raise0.2em\hbox{$>$}}}
 \def\ksim{\mathrel{\rlap{\lower0.2em\hbox{$\sim$}}\raise0.2em\hbox{$<$}}}
  \def\kg{\mathrel{\rlap{\lower0.25em\hbox{$>$}}\raise0.25em\hbox{$<$}}}

 \newcommand{\eq}[1]{(\ref{#1})}

\begin{document}

\title{Toward an understanding of the RHIC single electron data}

\author{P.B. Gossiaux and J. Aichelin}
\affiliation{SUBATECH, Universit\'e de Nantes, EMN, IN2P3/CNRS
\\ 4 rue Alfred Kastler, 44307 Nantes cedex 3, France}

\date{\today}

\begin{abstract} \noindent
High transverse momentum ($p_T$) single non-photonic electrons
which have been measured in the RHIC experiments come
dominantly from heavy meson decay. The ratio of their $p_T$ spectra
in pp and AA collisions ($R_{AA}(p_T)$) reveals the energy loss of
heavy quarks in the environment created by AA collisions. Using a
fixed coupling constant and the Debye mass ($m_D\approx gT$) as
infrared regulator perturbative QCD (pQCD) calculations are not able
to reproduce the data, neither the energy loss nor the azimuthal
$(v_2)$ distribution. Employing a running coupling constant and
replacing the Debye mass by a more realistic hard thermal loop (HTL)
calculation we find a substantial increase of the collisional energy
loss which brings the $v_2(p_T)$ distribution as well as
$R_{AA}(p_T)$ to values close to the experimental ones without
excluding a contribution from radiative energy loss.
\end{abstract}


\maketitle

\section{Introduction}
The spectra of mesons and baryons which contain light flavors
(u,d,s) only and which have been produced in ultrarelativistic heavy
ion collisions at the RHIC accelerator show a remarkable degree of
thermalization. Hydrodynamical calculations reproduce quantitatively
many of their dynamical properties and their multiplicity is well
described in statistical model calculations. Statistical
equilibrium, however, means loss of memory and therefore they are of
limited use for the study of the properties of the matter which is
created in the early phase of the reaction.

Heavy quarks, on the contrary, do not come to an equilibrium with
the surrounding matter and may therefore play an important role in
the search for the properties of this matter. Produced in hard
collisions, their initial momentum distribution can be directly
inferred from pp collisions. The deviation of the measured heavy
meson $p_T$ distribution in AA collisions (divided by $N_c$, the
number of binary initial collisions) from that measured in pp collisions,
is usually quantified as  $R_{AA}=d\sigma_{AA}/(N_c\
dp_T^2)/(d\sigma_{pp}/dp_T^2)$. $R_{AA}$ is a direct measure of the
interaction of the heavy quarks with the environment which is
created in AA collisions. The same is true for the azimuthal distribution,
$d\sigma/d\phi \propto (1+2 v_1\cdot \cos(\phi) + 2 v_2\cdot
\cos(2\phi))$, where the $v_2$ parameter is referred to as
``elliptic flow'', because at production no azimuthal direction is
preferred. The observed finite $v_2$ value is therefore either due
to interactions with light quarks and gluons or due to coalescence
at the end of the deconfined phase when the heavy quarks are
reshuffled into heavy mesons.

In the RHIC experiments heavy mesons have not yet directly been
measured. Both, the STAR \cite{Abelev:2006db} and the PHENIX
\cite{Adare:2006nq} collaboration, observe single non-photonic
electrons only. They have been created in the semileptonic decay of
heavy mesons. Thus experimentally one cannot separate between charm
and bottom hadrons. pQCD calculations in Fixed Order + Next to
Leading Logarithm (FONLL) predict a ratio of
$\sigma_{\bar{b}b}/\sigma_{\bar{c}c} = 7 x 10^{-3}$ with the
consequence that above $p_T > p_{T\,{\rm cross}} \approx 4\, {\rm
GeV}$ electrons from bottom mesons dominate the spectrum
\cite{Cacciari:2005rk}. The uncertainty of this value is, however,
considerable. The little known form of the electron spectrum from
heavy meson decay and the little known ratio of heavy quark mesons
to heavy quark baryons \cite{MartinezGarcia:2007hf} add to this
uncertainty.

In order to understand the single non-photonic electron spectra one has to
meet two challenges: One has to understand the interaction of a
heavy quark with the environment, produced in heavy ion collisions,
and one has to understand how this environment changes as a function
of time. In the past, several theoretical approaches
 \cite{Moore:2004tg,vanHees:2004gq,van
Hees:2005wb,Greco:2007sz,Svetitsky:1987gq,Gossiaux:2004qw,Gossiaux:2006yu,Molnar:2004ph,zhang,Armesto:2005mz}.
have been advanced to meet these challenges. Almost all of them
assume that in the heavy ion reaction a quark-gluon plasma (QGP) is
created and that the time evolution of the heavy quark distribution
function, $f(\vec{p},t)$, in the QGP can be described by a
Fokker-Planck approach  \be \frac{\partial f(\vec{p},t)}{\partial t}
=\frac{\partial}{\partial p_i}[ A_i(\vec{p})f(\vec{p},t)+
\frac{\partial}{\partial p_j}B_{ij}(\vec{p})f(\vec{p},t)]. \ee In
this approach the interaction of a heavy quark with the QGP is
expressed by a drag $(A_i)=<(p-p')_i> $ and by a diffusion
$(B_{ij}=\frac{1}{2}<(p-p')_i(p-p')_j>)$ coefficient calculated from
the microscopic $2\rightarrow 2$ processes by \bea
 <X> &=& \frac{1}{2E}
    \int \frac{d^3k}{(2\pi)^3 2k}
    \int \frac{d^3k'}{(2\pi)^3 2k'}\,
    \int \frac{d^3p'}{(2\pi)^3 2E'}
    \nonumber \\
    && \hskip -7mm n_i(k)
    \times\, (2\pi)^4\delta^{(4)}(p\!+\!k\!-\!p'\!-\!k')\frac{1}{d_i}
\sum \l|{\cal M}_i\r|^2\, X \, . \label{eq:meanv} \eea $p(p')$ and
$E=p_0$ ($E'=p_0')$ are momentum and energy of the heavy quark
before (after) the collision and k(k') is that of the colliding
light quark or gluon. $d_i$ is 4 for qQ and 2 for gQ. $n(k)$ is the
thermal distribution of the light quarks or gluons which is usually
taken as of Boltzmann type. ${\cal M}_i$ is the matrix element for
the channel i, calculated using pQCD Born matrix elements. Up to now
the calculations are limited to elastic collisions (Qq and Qg). The
matrix elements for these channels can be found in ref.
\cite{Svetitsky:1987gq,Combridge:1978kx}. They contain 2 parameters
which have to be fixed: the coupling constant and the infrared (IR)
regulator to render the cross section infrared finite. Up to now all
calculations have used a fixed coupling constant, albeit different
numerical values. As IR regulator usually a Debye mass $m_D$ has
been employed which is assumed to be proportional to the thermal
gluon mass $m_D = \beta gT$ with $\beta$ around 1.

The Fokker-Planck approaches differ in the way in which the
surrounding matter is taken into account. The Texas A\&M group
\cite{vanHees:2004gq,van Hees:2005wb,Greco:2007sz} uses an expanding
fireball whereas the other groups
\cite{Moore:2004tg,Gossiaux:2004qw,Gossiaux:2006yu} use
hydrodynamical calculations, with different equations of state,
however.

Despite of different choices for \as and $m_D$  and of the different
models for the expansion of the QGP it is a common result of all of
these approaches that they underpredict by far the modification
of the heavy quark distribution due to the QGP. One has to
multiply the pQCD cross sections artificially by a K factor of the
order of $K \approx 10$ (which depends on the choice of \as and of
the IR regulator) to obtain agreement with experimentally
observed values for $R_{AA}(p_T)$ and for $v_2(p_T)$
\cite{Moore:2004tg,Gossiaux:2004qw,Gossiaux:2006yu}.

One possibility to reduce the value of K has been advanced by van
Hees et al. \cite{vanHees:2004gq,van Hees:2005wb} who assumed that
heavy D-mesons can be formed in the plasma and decay thereafter
isotropically. One has, however, to await more precise lattice
results to see whether such a nonperturbative process is indeed
possible.

It is the purpose of this article to improve these models in three
directions: 1) we replace the Fokker-Planck equation by a Boltzmann
equation because the momentum transfer is not well parameterized by
the first and second moment only. 2) we introduce a physical running
coupling constant, fixed by the analysis of $e^+e^-$ annihilation
and of the $\tau$ decay, in the pQCD matrix elements. 3) we replace
the ad hoc parametrization of the infrared regulator by one which
yields the same energy loss as the HTL energy loss calculations
\cite{Chakraborty:2006db,Mustafa:2004dr}. We will show that with
these new ingredients pQCD calculations yield a larger stopping of
heavy quarks in matter and bring the results of the calculation
close to the experimental values of $R_{AA}(p_T)$ and of $v_2(p_T)$.

We do not address here the radiative energy loss whose importance is
highly debated \cite{Moore:2004tg,Djordjevic:2003zk,Zakharov:2007pj}
because detailed microscopic calculations are not at hand yet. They
may easily count for the factor of two which remains for
$R_{AA}(p_T)$ between the data and the calculation which includes
collisional energy loss only. This will be the topic of an upcoming
publication.

\section{Infrared regulator}
In order to calculate the drag and diffusion coefficients (eq.
\ref{eq:meanv}) using pQCD Born matrix elements
\cite{Svetitsky:1987gq,Combridge:1978kx} the gluon propagator in the t-channel
has to be IR regulated by a screening mass $\mu$ \be
    \frac\alpha t \to \frac{\alpha}{t-\mu^2} \, .
\label{aborn} \ee Frequently the IR regulator is taken as the
thermal gluon mass
 \cite{Weldon:1982aq} \be \mu^2= \frac {m_D^2}{3}
    = \frac{N_c}{9} \l( 1+\T\frac16\, n_f \r) 4 \pi \, \alpha_S \, T^2\approx
\frac{(g_ST)^2}{3}\,
    \label{eq: mg}\ee
where $n_f\, (N_c)$ are the number of flavors (colors) and $m_D$ is
the Debye mass. The infrared regulator is, however, not very well
determined on first principles. Therefore, in the actual
calculations \cite{vanHees:2004gq,van
Hees:2005wb,Greco:2007sz,Svetitsky:1987gq} $\mu^2$ was taken in
between $g_S^2\,T^2$ and $\frac{g_S^2\,T^2}{3}$ with $g_S^2
=4\pi$\as. The IR regulator is one of the main sources of
uncertainty for the determination of the cross section (and hence
for the drag and the diffusion coefficient) and it is therefore
useful to improve its determination by physical arguments.

For QED Braaten and Thoma \cite{Braaten:1991jj} have shown that in a
medium with finite temperature the Born approximation is not
appropriate for low momentum transfer $|t|$. It has to be replaced
by a hard thermal loop (HTL) approach to the gluon propagator. At
high $|t|$ we can use the bare gluon propagator (left hand side of
eq. \ref{aborn}). This approach we call HTL + hard calculation. To
calculate differential cross sections using hard thermal loops is
beyond present possibilities but Braaten and Thoma have shown that
in QED the energy loss can be calculate analytically in the HTL +
hard approach. Our strategy is now the following: We assume that the
gluon propagator can be written in the form
\begin{equation}
%\frac\alpha t \to
\frac{\alpha}{t-\kappa m_D^2(T)}\,,
\label{eq:eff_prop_const}
\end{equation}
and determine the value of $\kappa$ by requiring that a pQCD Born
calculation with this gluon propagator gives the same energy loss as
the HTL + hard approach.

We first deal with the QED case where the underlying hypothesis $g^2
T^2 \ll T^2$ is more likely to be satisfied and focus our attention
on the $t$-channel which is the only one suffering from IR
singularities and therefore decisive for the choice of $\kappa$. For
the HTL + hard approach we follow ref.
\cite{Braaten:1991jj,Peigne:2007sd} where the collision of a muon
with an electron is calculated. Let us consider the energy loss \bea
-\frac{dE_\mu}{dx}
    &=&\frac{1}{2E v}
    \int \frac{d^3k}{(2\pi)^3 2k}
    \int \frac{d^3k'}{(2\pi)^3 2k'}\,
    \int \frac{d^3p'}{(2\pi)^3 2E'}
    \nonumber \\
    && \hskip -7mm n_F(k)(1-n_F(k'))
    \times\, (2\pi)^4\delta^{(4)}(p\!+\!k\!-\!p'\!-\!k') \frac{1}{d}\sum \l|{\cal M}_{\mu e\rightarrow \mu' e'}\r|^2\, \omega \,,
\label{eq:eloss} \eea where $v$ is the velocity of the heavy muon,
$\omega = E-E'$ is the energy transfer in the collision and $d = 4$
is the overall spin degeneracy. The total energy loss is the sum of
two contributions:
\begin{enumerate}
\item
At small momentum transfer $|t|=|(p-p')^2|<|t^*|$, where $|t^*|$ is
an intermediate scale chosen between $g^2T^2$ and $T^2$, the hard
thermal loop regulates the infrared singularity and we obtain
(\cite{Braaten:1991jj,Peigne:2007sd})\be -\l. \frac{dE_\mu}{dx}
\r|_{|t| < |t^\star|}^{v \to 1} = \frac{g^4T^2}{48\pi}\, \ln{\frac{6
|t^\star|}{g^2 T^2}} \, .\label{eq:elosssoft} \ee
\item
At large $|t|$ ($|t|_{\rm max}>|t|> |t^*|$) no infrared regulator is
necessary and we arrive at
(\cite{Braaten:1991jj,Peigne:2007sd}) \bea
-\left.\frac{dE_\mu}{dx}\right|^{v\rightarrow 1}_{|t|>|t^*|} \approx
\frac{g^4 T^2}{48 \pi}\left[
\ln\frac{8ET}{|t^*|}-\gamma-\frac{3}{4}-\frac{\zeta'(2)}{\zeta(2)}\right]\,
. \label{hardPP2} \eea
\end{enumerate}
Adding the HTL (eq. \ref{eq:elosssoft}) and the hard (eq.
\ref{hardPP2}) part, the intermediate scale $t^*$ disappears and we
arrive at \cite{PP}
\begin{equation}
-\left.\frac{dE_\mu}{dx}\right|^{v\rightarrow 1}_{\rm HTL+hard}\approx
\frac{g^4 T^2}{48 \pi}\left[\ln\frac{48ET}{g^2 T^2}-\gamma-\frac{3}{4} -
\frac{\zeta'(2)}{\zeta(2)}\right]\,.
\label{dedxPP}
\end{equation}
We compare now this result with that obtained by introducing an
infrared regulated gluon propagator eq. \ref{aborn}. In Born
approximation we obtain the cross section:
\begin{equation}
\frac{d\sigma_F}{d t}= \frac{g^4}{\pi(s-M^2)^2}\,
\left[\frac{(s-M^2)^2}{(t-\mu^2)^2}+\frac{s}{t-\mu^2}+
\frac{1}{2}\right]\,. \label{eq:born}
\end{equation}
We evaluate here the energy loss for the whole t-interval $t\in
[t_{\rm min},0]$ and obtain (for details we refer to the appendix A)
\begin{equation}
-\left.\frac{dE_\mu}{dx}\right|^{v\rightarrow 1}_{\rm eff}\approx
\frac{g^4 T^2}{48 \pi}\left[ \ln\frac{8ET}{e
\mu^2}-\gamma-\frac{3}{4}-\frac{\zeta'(2)}{\zeta(2)}\right]\,.
\label{dedemod2}
\end{equation}
Comparing the pQCD Born (eq.\ref{dedemod2}) with the HTL + hard
result  (eq. \ref{dedxPP}), we find that $\mu^2$ has to be
\begin{equation}
\mu^2=\frac{g^2T^2}{6e}=\frac{3}{2e}m_{\gamma}^2=
\frac{m_D^2}{2e}\quad\Rightarrow\quad
\kappa=\frac{1}{2e}\approx 0.2\,.
\label{eq:optkappa_qed}
\end{equation}
in order to obtain the same energy loss in QED.

\begin{figure}[H]
\epsfig{file=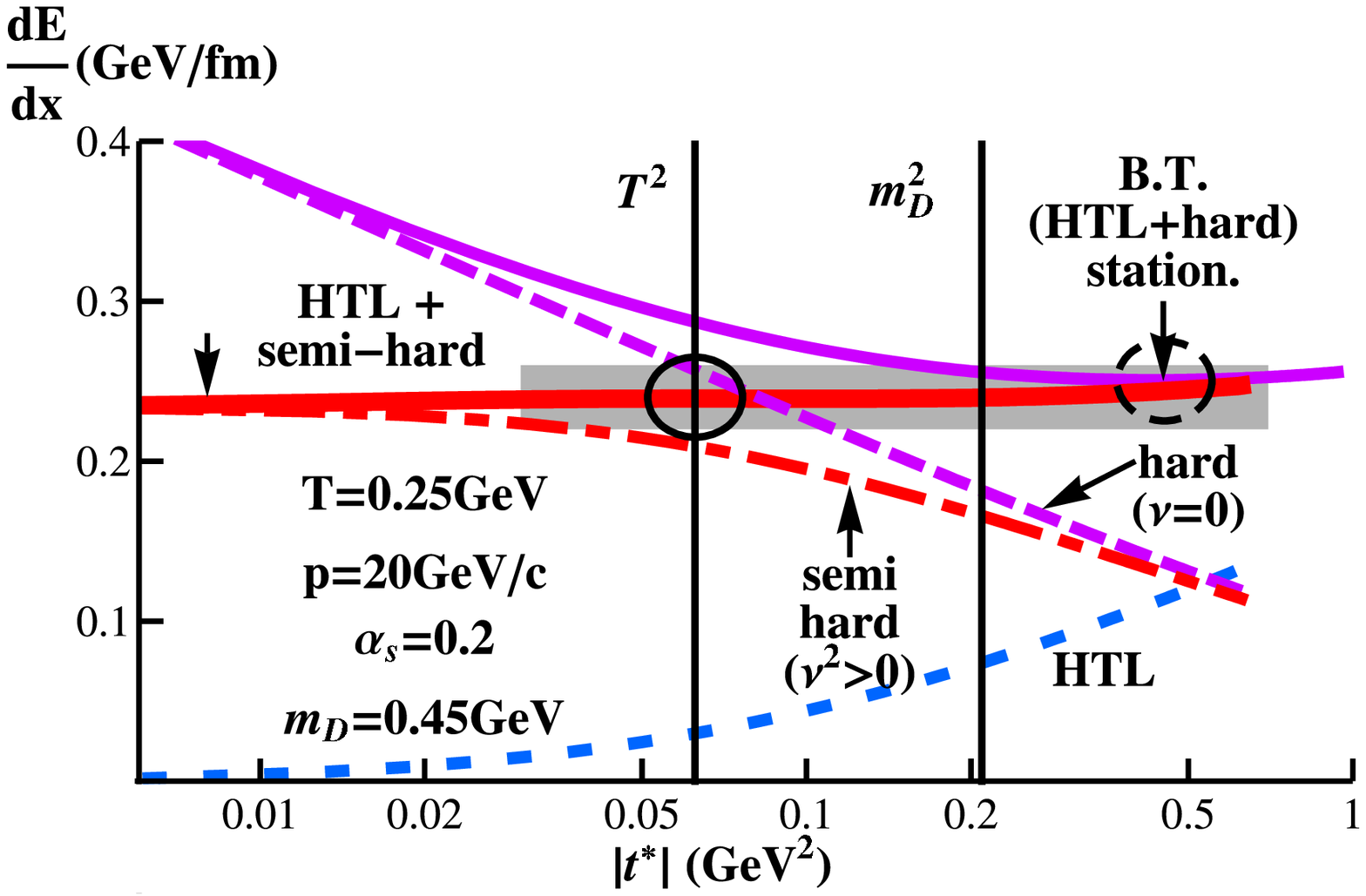,width=0.49\textwidth}
\epsfig{file=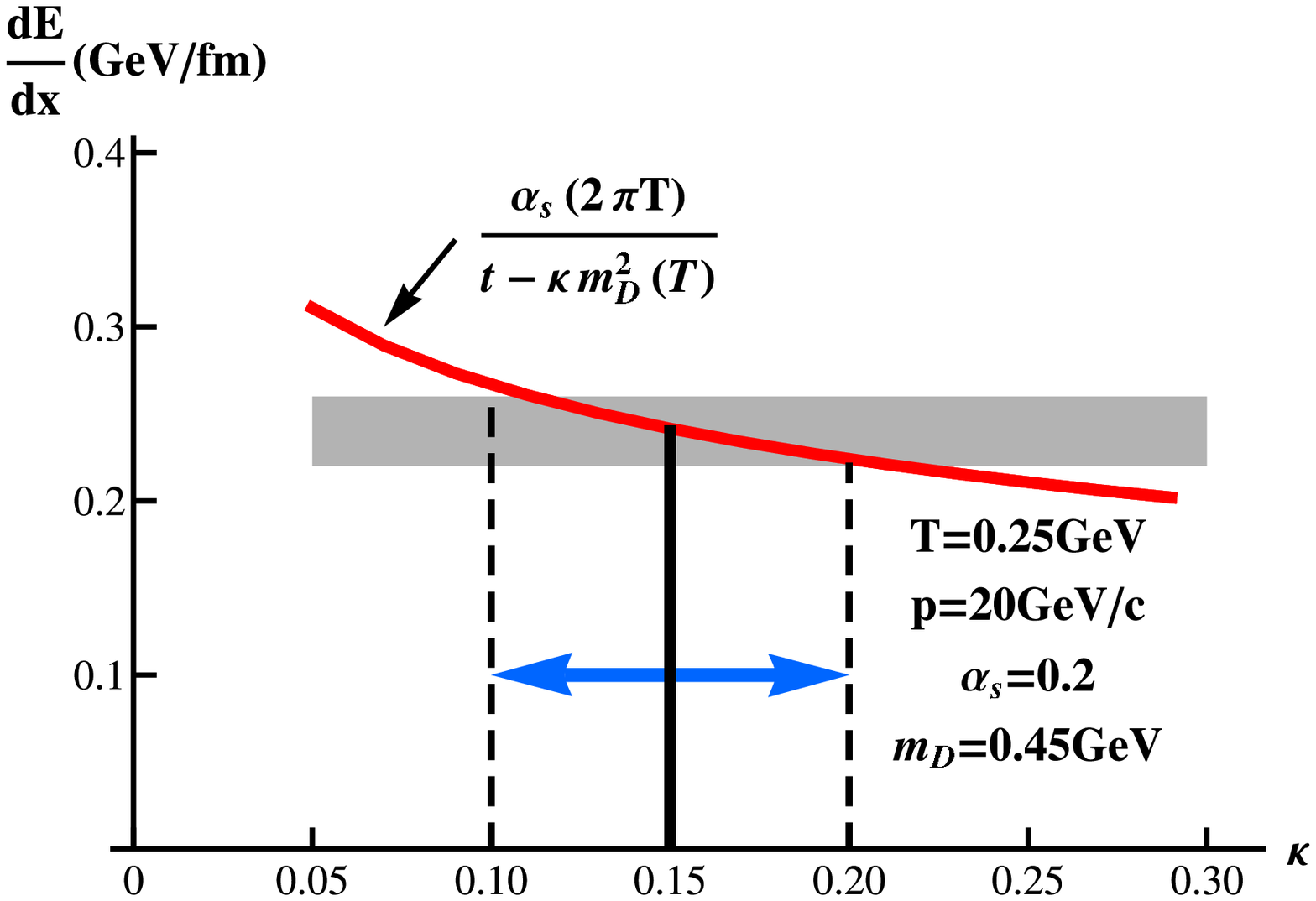,width=0.49\textwidth} \caption{Left:
the total energy loss in the HTL + hard approach as well as the
different components  for a given choice of parameters as a function
of the intermediate scale $t^*$. The full lines are the sum of the
HTL (blue dotted) and hard/semi-hard part (dashed purple for $\nu^2
= 0$, dashed-dotted red for $\nu^2 \approx 0.16 m_D^2$). Right: The
total energy loss evaluated with Born cross sections and with the
propagator, eq. \ref{eq:eff_prop_const}, as a function of $\kappa$.
Only $t$-channel contribution has been considered here.}
\label{arun}
\end{figure}

Because QED and QCD have a very similar HTL-propagator structure the
above approach remains valid for QCD as well provided that $\alpha_S
\ll 1$ and that $\mu^2$ is replaced by eq. \ref{eq: mg}.
% previous version
% \textcolor{blue}{In the QCD case there is, however, the complication
% that in the range of validity of the HTL approach, $m_D^2\ll T^2$ ,
% the HTL+hard model is not independent on the intermediate scale
% $t^*$ for temperatures obtained at RHIC.}
% PBG proposal:
In the QCD case there is, however, the complication that we are, for
temperatures achieved at RHIC, at the best at the borderline of the
the range of validity of the HTL approach, $m_D^2\ll T^2$. As a
consequence, the HTL+hard model -- commonly used by many authors --
is in fact not independent on the intermediate scale $t^*$. To
demonstrate this problem we start out as in QED. For small $|t|$ we
obtain
\begin{equation}
-\left.\frac{dE_Q}{dx}\right|_{|t|<|t^*|}=\frac{C_F \alpha_S}{v^2}
\int_{-v}^{v}
\frac{x}{(1-x^2)^2}\int_{t^*}^{0}dt (-t)\left[\rho_L+(v^2-x^2)\rho_T\right]
\label{eq:htl_qcd}
\end{equation}
with $v$ being the velocity of the heavy quark $Q$ and the spectral functions
\begin{equation}
\rho_L(t,x)\equiv
-\frac{1}{\pi}\Im\left[\frac{1}{\frac{-t}{1-x^2}+\Pi_L(x)}\right]\quad
\text{and}\quad \rho_L(t,x)\equiv
-\frac{1}{\pi}\Im\left[\frac{1}{t+\Pi_T(x)}\right].\,
\end{equation}
$\Pi_L$ and $\Pi_T$ are the self-energies evaluated in the HTL
approximation:
\begin{eqnarray}
\Pi_L(x)&=&m_D^2\left(1-\frac{1}{2}\ln\left|\frac{1+x}{1-x}\right|+
\frac{i\pi x}{2}\right)\nonumber\\
\Pi_T(x)&=&\frac{m_D^2}{2}\left(x^2+\frac{x(1-x^2)}{2}
\ln\left|\frac{1+x}{1-x}\right|+\frac{i\pi x(x^2-1)}{2}\right).\,
\label{htl}
\end{eqnarray}
For large $|t|$ we obtain (see eq. \ref{eq:eloss})
%$-\left.\frac{dE_Q}{dx}\right|_{|t|>|t^*|}$
\bea -\left.\frac{dE_Q}{dx}\right|_{|t|>|t^*|}
    &=&\sum_i\frac{1}{2Ev}
    \int \frac{d^3k}{(2\pi)^3 2k}
    \int \frac{d^3k'}{(2\pi)^3 2k'}\,
    \int \frac{d^3p'}{(2\pi)^3 2E'}\,\Theta(|t|-|t^*|)
    \nonumber \\
    && \hskip -7mm n_i(k)(1\mp n_i(k'))
    \times\, (2\pi)^4\delta^{(4)}(p\!+\!k\!-\!p'\!-\!k') \frac{1}{d_i}
\sum \l|{\cal M}_{i}\r|^2\, \omega .\, \label{eq:eloss_qcd_hard}
\eea Here the matrix elements include $qQ\rightarrow qQ$ as well as
$gQ\rightarrow gQ$ collisions.  In contradistinction to the QED case
the sum of both terms depends explicitly on the intermediate scale
$t^*$ in the region $[m_D^2,T^2]$, as seen in fig. \ref{arun}.
There we display the two parts of the energy loss (eq.
\ref{eq:htl_qcd}, blue dotted and eq. \ref{eq:eloss_qcd_hard},
purple dashed) as well as the sum of both (purple full). Clearly,
the total energy loss becomes stationary with respect to the
intermediate scale $|t^*|$ only for a value of $|t^*|\approx
0.4\,{\rm GeV}^2\gg T^2 (=0.0625\,{\rm GeV}^2)$ and hence in a
region where the HTL approach is not valid anymore. Mathematically,
this is due to the appearance of terms $\propto
O(\frac{m_D^2}{|t^*|})$ and $\propto O(\frac{|t^*|}{T^2})$ which are
neither small nor do they compensate. Physically, we are in a regime
where the interaction is screened over a distance of the same order
as the mean distance between QGP constituents, so that a large part
of the ``hard collisions'' will be affected by the medium
polarization as well. Our prescription to cure this problem is to
add an IR regulator $\nu^2$ to the hard part (as $\mu^2$ in eq.
\ref{aborn}). We dubbed this approach therefore 'semi-hard'. The HTL
part remains unchanged. The value of $\nu^2$ is chosen in that way
that for a wide range of temperatures and heavy-quark momenta the
sum of the HTL and semi-hard energy loss is independent of $t^\star$
for $|t^\star|<T^2$ i.e. in the range where the HTL approximation
holds. The red bold line in figure \ref{arun} shows this
independence of the total energy loss on  $t^\star$ when the hard
part is replaced by the semi-hard (red dashed dotted) approach for
$p=20\,{\rm GeV}$, $T=0.25\,{\rm GeV}$ and $\nu^2\approx 0.16
m_D^2$. We will adopt this value of $\nu^2$ for the further
calculations.

If we compare the t channel energy loss calculated in the HTL +
semi-hard approach (shaded area in figure \ref{arun}) with that
obtained within our pQCD Born approach (eq. \ref{eq:eff_prop_const})
we find a value of $\kappa$ around 0.15. This value is close to that
obtained in QED (eq.\ref{eq:optkappa_qed}). It is considerably lower
than those used up to now in the pQCD cross section calculation.
This is our first seminal result.

\section{Running coupling constant}
The constant coupling constant \as is the other quantity which
limits the predictive power of the present calculations. In the
published calculations \as was taken in between 0.2
\cite{Braaten:1991jj} and 0.6 \cite{Svetitsky:1987gq} leading to a
difference of a factor 9 for the drag and for the diffusion
coefficient.

\begin{figure}[h]
\epsfig{file=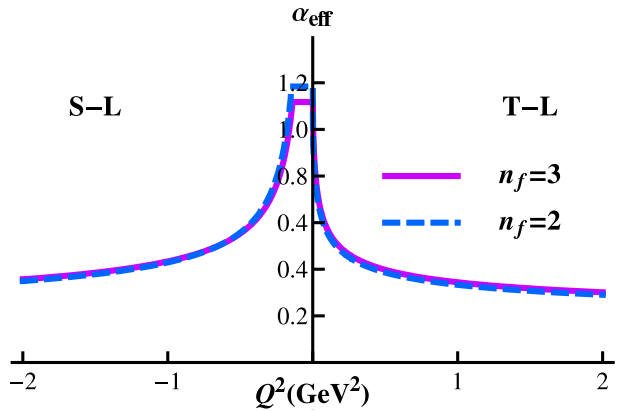,width=0.49\textwidth}
\caption{$Q^2$ dependence of the running coupling constant.}
\label{arunb}
\end{figure}

As has been observed by Dokshitzer \cite{Dokshitzer:1995qm} there
exists the possibility to define a running coupling which stays
finite in the infrared by writing observables as a product of an
universal effective time-like coupling and a process dependent
integral. An alternative approach is to define an effective coupling
constant, $\alpha_{\rm eff}(Q^2)$, from the analysis of physical
observables. Two different experiments, $e^+e^-$ annihilation
\cite{Mattingly:1993ej} as well as non-strange hadronic decays of
$\tau$ leptons \cite{Brodsky:2002nb}, have been used to determine
the infrared behavior of $\alpha_{\rm eff}(Q^2)$ . The resulting
coupling constants are infrared finite and very similar. These
effective couplings are all-order resummations of perturbation
theory and include all non-perturbative effects. We extend the
parametrization of the time-like sector,
Ref.~\cite{Dokshitzer:1995qm}, to the space like sector, which leads
to
 \be
   \alpha \to \alpha_{\rm eff}(Q^2)
   =
  \frac{4\pi}{\beta_0}
   \l\{ \begin{array}{lc}
   L_-^{-1}
   \\[-0.35em]
   &   \ {\rm for \ } Q^2 \kg 0 \, ,
   \\[-0.35em]
   \frac12 - \pi^{-1} {\rm atn}( L_+/\pi )
   \end{array} \r.
   \label{eq: alpha(Q2)}
\ee
with $\beta_0 = 11-\frac23\, \nf$, $\nf=3$, and
$L_\pm = \ln(\pm Q^2/\Lambda^2)$ . In the space-like
sector we replace the propagator
 \be
\frac\alpha t \to \frac{\alpha_{\rm eff}(t)}{t-\mu^2} \, ,
\label{eq:dm} \ee where $\mu^2$ is an IR regulator which we will
specify below. The coupling constant $\alpha_{\rm eff}$ is displayed
in fig. \ref{arunb} for 2 and for 3 flavors.
It has already  been argued in \cite{Peshier:2006ah} that a running
coupling constant leads to the disappearance of the logarithmic E
dependence of the energy loss at large energies:
\begin{equation}
\frac{dE}{dx}\propto \alpha_S(2\pi T)^2 T^2 \ln\frac{E T}{m_D^2}
\quad\longrightarrow \quad \frac{dE}{dx}\propto \alpha_S(\mu^2)T^2\,,
\label{eq:behav_eloss}
\end{equation}
with an IR regulator $\mu^2=[\frac{1}{2},2]\,\tilde{m}_D^2$, where
the Debye mass $\tilde{m}_D$ is determined self-consistently
according to \be \mds^2(T)= \frac{N_c}{3} \l( 1+\T\frac16\, \nf
\r)4\pi\,\alpha(-\mds^2(T)) \,T^2\,. \label{eq:md} \ee However, this
ambiguity of the coefficient leads to a non negligible uncertainty
in the energy loss.

\begin{figure}[H]
\epsfig{file=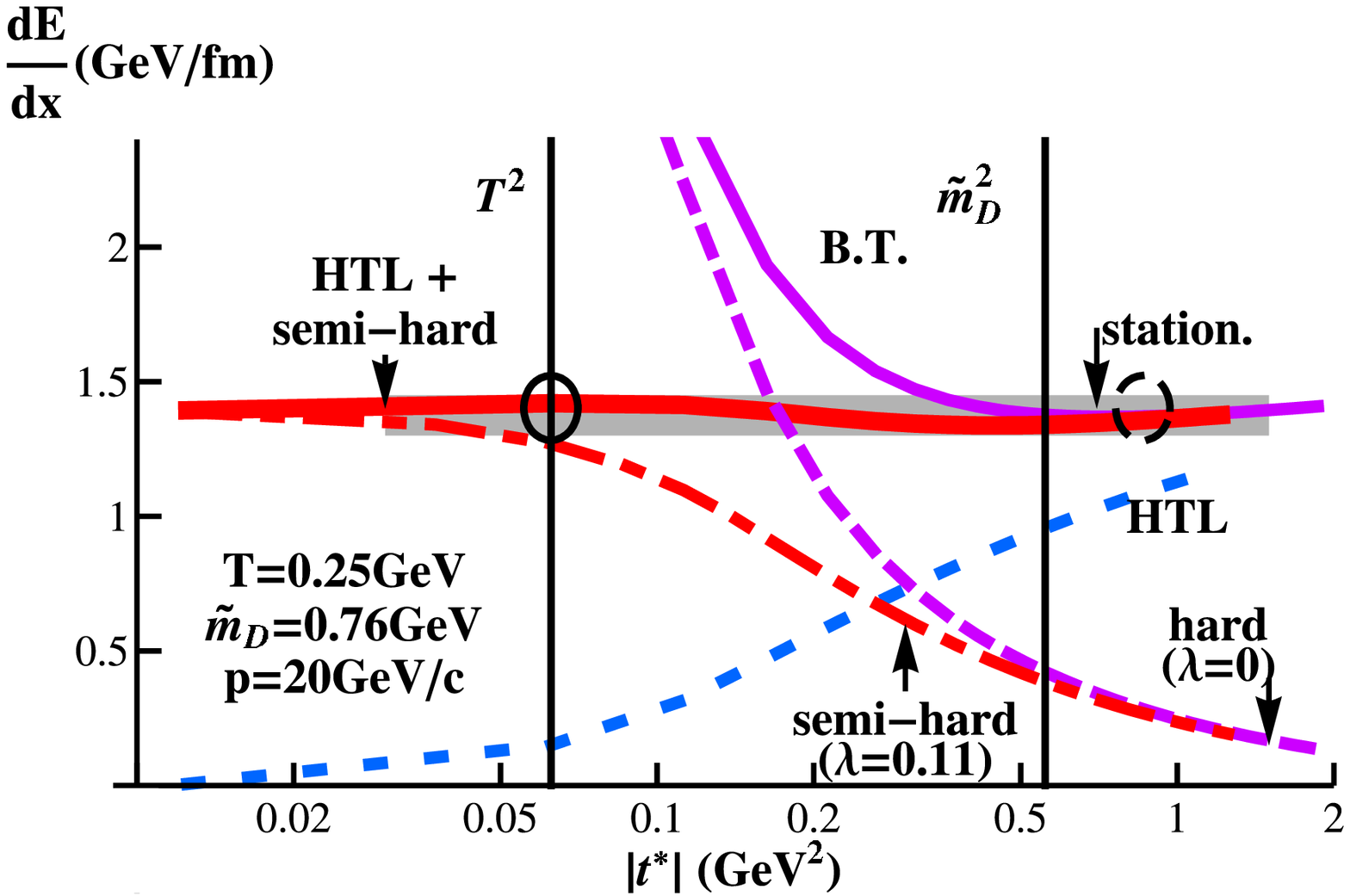,width=0.49\textwidth}
\epsfig{file=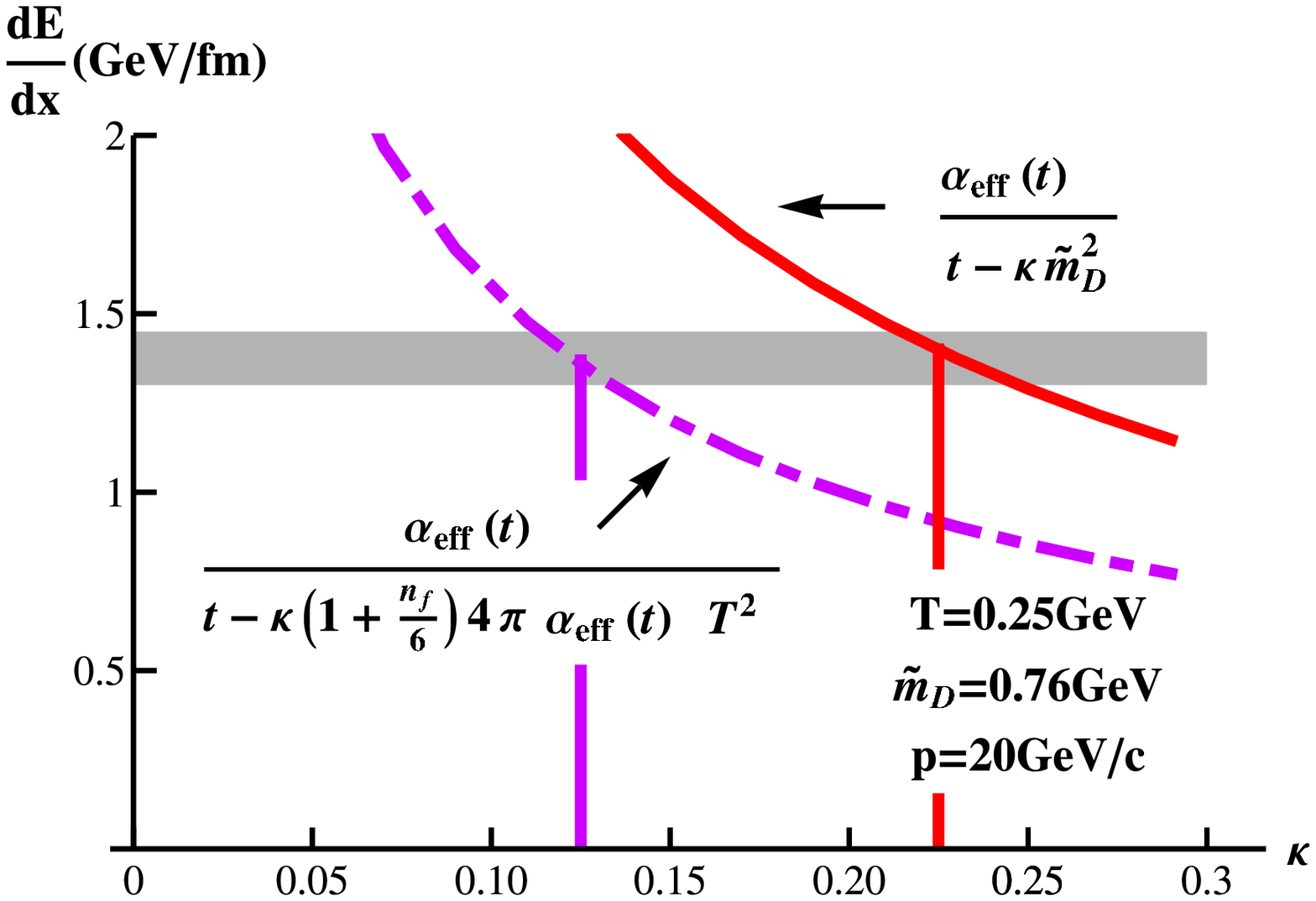,width=0.49\textwidth}
\caption{Left: Same quantities as in fig. \ref{arun} for the case of a running
$\alpha_{\rm eff}$. Right: The total energy loss in pQCD Born
approximation for two different infrared regulators as a function of
$\kappa$. The shaded area corresponds to the energy loss calculated
in the HTL + semi-hard approach (left).}
\label{arunc}
\end{figure}

In this work, we determine the optimal infrared regulator using the
same strategy as for the non-running case: we calibrate the energy
loss to the one obtained in a generalized  ``HTL + semi-hard''
approach this time with a running coupling constant. For this
purpose, we assume that the (squared) Debye mass for a fixed
coupling constant appearing in the hard thermal loop terms (eq.
\ref{htl}), $m_D^2(T)\equiv (1+\frac{n_f}{6})g^2 T^2$, can be
replaced by $m_D^2(T,t)\equiv (1+\frac{n_f}{6}) 4\pi \alpha_{\rm
eff}(t) T^2$ . As illustrated on fig. \ref{arunc} (left, full purple
line), also here  the total energy loss depends on the intermediate
scale $|t^\star|$ in the domain of validity of the HTL approach, if
we employ the HTL+hard approach. Only if we replace the hard by a
semi-hard propagator
\begin{equation}
\frac{\alpha_{\rm eff}(t)}{t}
\quad \longrightarrow \quad
\frac{\alpha_{\rm eff}(t)}{t-\lambda m_D^2(T,t)}\,,
\label{eq:prop_run_model}
\end{equation}
we may obtain an energy loss which is independent on the intermediate scale
$t^\star$. The optimal choice
is  $\lambda\approx 0.11$ (see fig. \ref{arunc}, left, bold red line).
Using this prescription, the energy loss in the $t$-channel is found
to be $\approx 1.3-1.4~{\rm GeV}/{\rm fm}$ i.e. $\approx 6$ times
larger than the energy loss found with the same parameters for the
non-running coupling constant. For $|t^\star|< T^2$, the HTL
contribution becomes negligible and the energy loss is given by the
semi-hard part  only (which is IR-convergent). Therefore, the
natural IR regulator $\mu^2$ for our effective Born pQCD approach
(eq. \ref{eq:dm}) is $\mu^2=\kappa m^2_D(T,t)$, with
$\kappa\approx\lambda\approx 0.11$, i.e. exactly the propagator of
the rhs of eq. \ref{eq:prop_run_model}.

However, the same energy loss can be obtained if one uses the
simpler propagator of eq. \ref{eq:dm} taking $\mu^2=\kappa
\tilde{m}_D^2(T)$ \be \frac{\alpha_{\rm eff}(t)}{t-\kappa
\mds^2(T)}\,, \label{eq:dmbis} \ee with $\kappa\approx 0.2$ and
$\mds$ the Debye mass defined self-consistently according to eq.
\ref{eq:md}. This is shown on the right hand side of fig.
\ref{arunc} and leads to our choice $\mu^2_{QCD}=0.2\,\mds^2(T)$ for
the propagator defined by eq. \ref{eq:dm}. We will show later that
with these values the drag coefficient and hence the energy loss
differs only slightly between these two models in the $(T,p)$ range
of interest for ultrarelativistic heavy ion collisions. We note in
passing that a similar energy loss has been obtained by Wick et al.
\cite{Wicks:2007mk} in a simpler model for light quarks.

\section{results}
\label{sect:results}

In order to evaluate the consequences of our new approach we compare
the results with those obtained for other choices of coupling
constants and infrared regulators. They are summarized in table 1.
From A $\rightarrow$ F the parameterizations become increasingly
realistic.
\begin{table}[h]
\begin{center}
\begin{tabular}{|c|c|c|c|c|}
\hline&\as & $\mu^2$&line form& figure color\\
\hline
A\ & 0.3 & $m_D^2$&dotted thin &black\\
B\ & $\alpha_S(2\pi T)$ & $m_D^2$&dashed thin&black\\
C\ & $\alpha_S(2\pi T)$ & $0.15 \times m_D^2$&full thin&black \\
D\ & running (eq.\ref{eq: alpha(Q2)}) & $\tilde{m}_D^2$&dashed bold&red\\
E\ & running (eq.\ref{eq: alpha(Q2)}) & $0.2 \times \tilde{m}_D^2$&full bold& red \\
F\ & running (eq.\ref{eq: alpha(Q2)}) &
$0.11\times 6\pi \,\alpha_{\rm eff}(t)\,T^2$ & dashed dotted bold& purple\\
\hline
\end{tabular}
\caption{Coupling constants and infrared regulators used in our
calculations}
\end{center}
\label{tabl}
\end{table}
\smallskip
For the results presented below we include the s and u channels as
well. They do not require any IR regulator and the coupling constant
have been chosen as $\alpha \rightarrow \alpha_{\rm eff}(s-m^2)$ and
$\alpha \rightarrow \alpha_{\rm eff}(u-m^2)$ because $s=m^2$ and
$u=m^2$ correspond to the maximal ``softness'' in these channels.

\subsection{Cross sections}
\begin{figure}
\epsfig{file=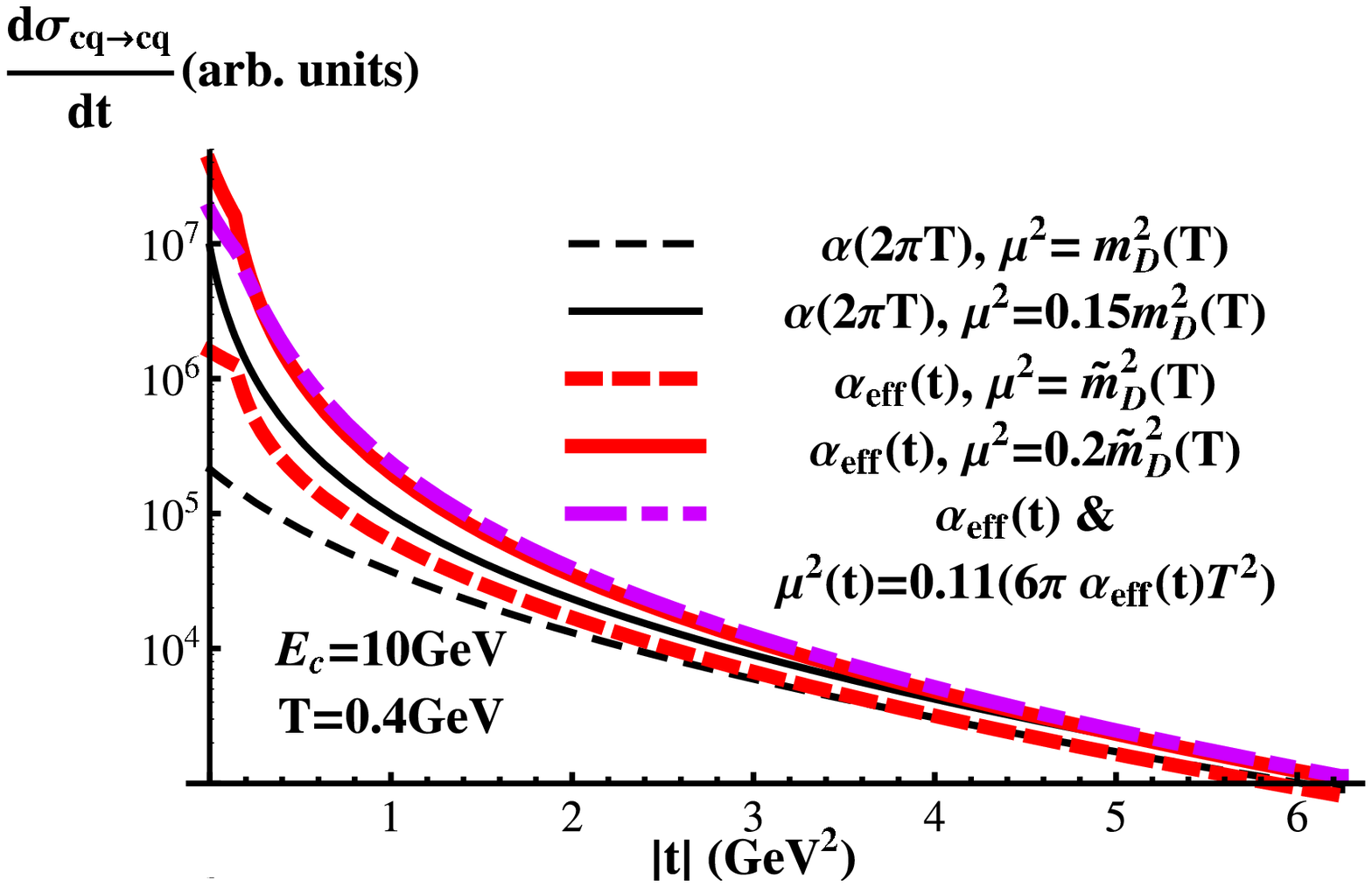,width=0.49\textwidth}
\epsfig{file=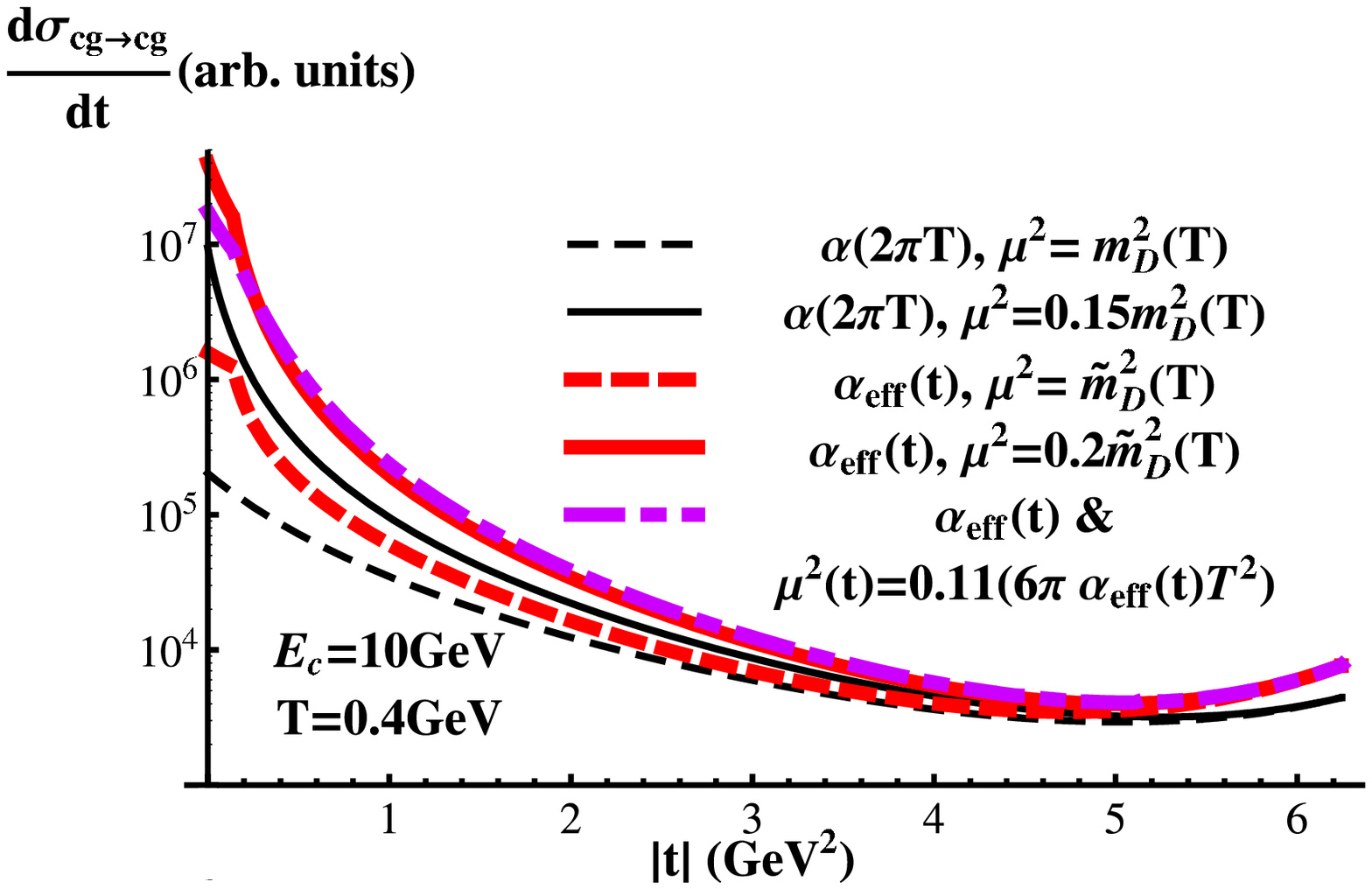,width=0.49\textwidth} \caption{Effective
cross section $d\sigma/dt$ for the different models (see table 1).}
\label{cross}
\end{figure}

The cross sections $\frac{d\sigma}{dt}$ for the different
parameterizations of table 1 are displayed in fig. \ref{cross}, left
for quarks and right for gluons.
It is evident that both, a running coupling constant and a lower IR
regulator, increase the cross section at small t whereas the
increase at high t is rather moderate, but nevertheless
visible in the $gQ$ reactions, due to the u-channel.

\subsection{Individual collisions and transport coefficients}

For many interpretations it is interesting to see how the quarks
loose their energy when traversing a plasma of a given temperature.
For this purpose we
study the differential probability $P_i(w,p)$ that a heavy quark
with a momentum $p$ in the rest system of the heat bath looses
the energy $w$ by colliding with a plasma particle of
type $i$:
\be
  P_i(w, p)
  \equiv
  \int\frac{d^3k}{(2\pi)^3} \frac{n_i(k)}{2k}
  \int_{t_-}^{t_+}\! \frac{dt}{\sqrt{H}}\, \sum\l|{\cal M}_i\r|^2.
\label{eq:P} \ee  The condition $H \ge 0$, where \bea
    H
    &=& (4\pi)^4 E^2 \l[
    \l(s-(E+k)^2 \r)t^2
    +
    \l( (2Ek-s+m_c^2)^2 \r. \r.
    \nonumber \\
    && \quad
    \l.\l. - 4k^2p^2+2w(k(s+m_c^2)-E (s-m_c^2))  \r)t- w^2 (s-m_c^2)^2 \r]\,,
\eea
with $s=m_c^2+2Ek(1-\cos\theta(\vec{k},\vec{p}))$, determines
not only the limits $t_\pm$ in eq.~\eq{eq:P}, it also constrains the
integral over $\vec{k}$.

\begin{figure}[htb]%[h]
\epsfig{file=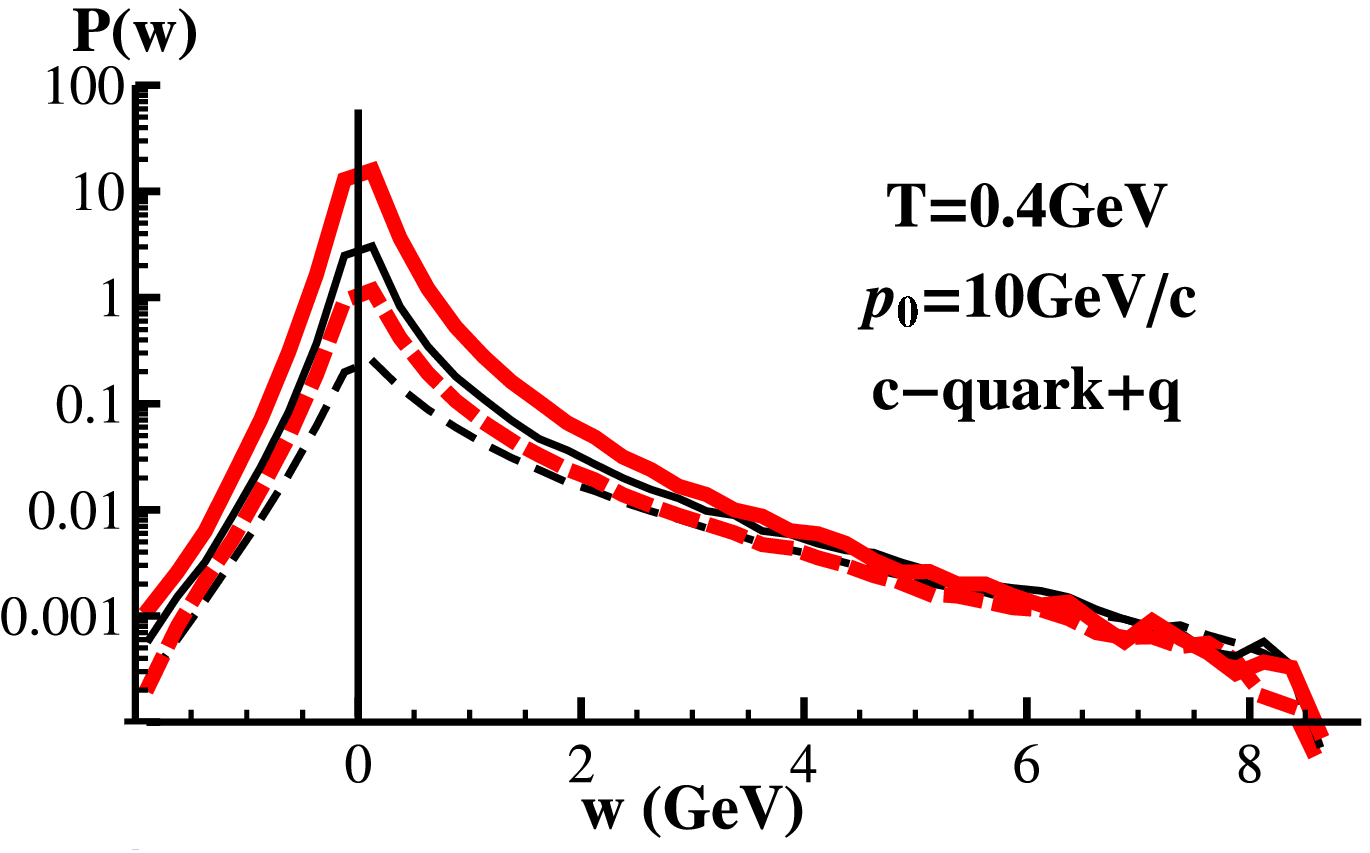,width=0.49\textwidth}
\epsfig{file=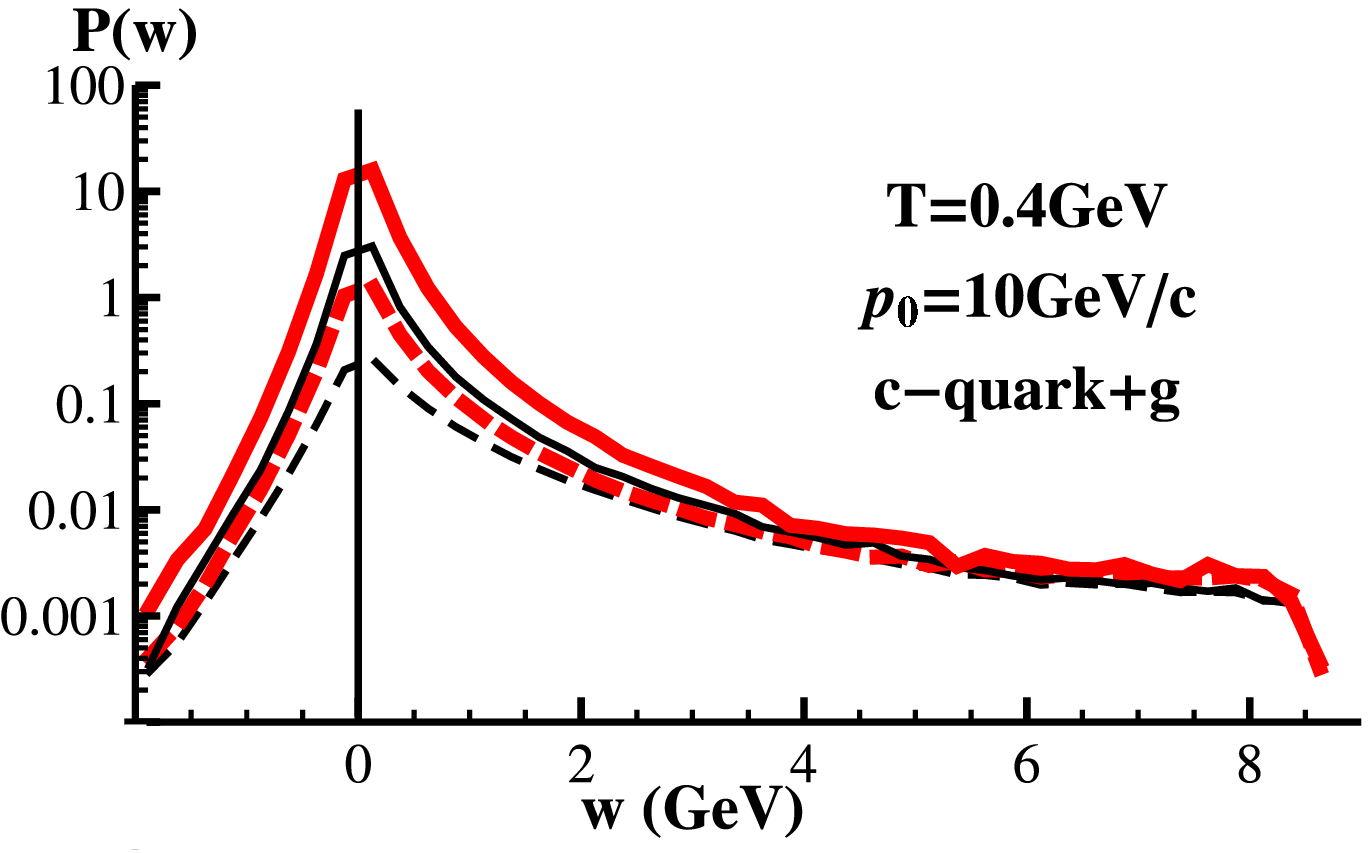,width=0.49\textwidth} \caption{Differential
probability $P(w)$ that a c-quarks with an initial momentum of p =
10 GeV/c looses the energy $w$ in a collision with a plasma particle
in a plasma at $T = 400$ MeV, on the left hand side for collisions
with quarks, on the right hand side for collisions with gluons. For
the different curves see table 1.} \label{pomega}
\end{figure}

The probability $P_i(w,p)$ for c-quark with $p = 10\,{\rm GeV}$ in a
plasma of the temperature of $T=400\,{\rm MeV}$ is displayed in
fig.\ref{pomega}. On the left (right) side we see the probability
for cq (cg) collisions. Negative values of $w$ mean that the heavy
quark gains energy in the collision. Due to the u-channel
contribution cg collisions are more effective to transfer a large
amount of energy. The large majority of the collisions yield only a
small energy transfer.
\begin{figure}[htb]%[h]
\epsfig{file=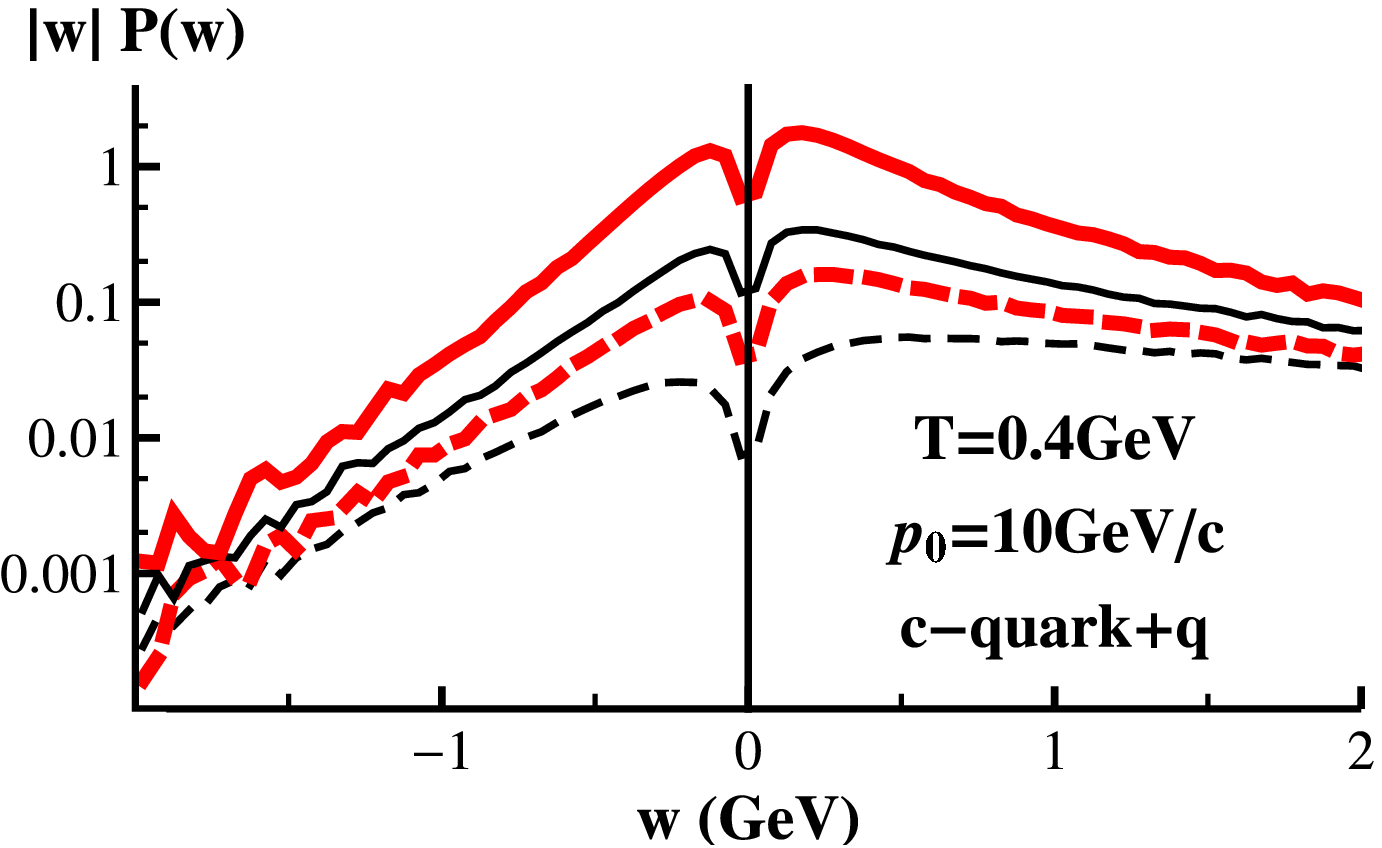,width=0.49\textwidth}
\epsfig{file=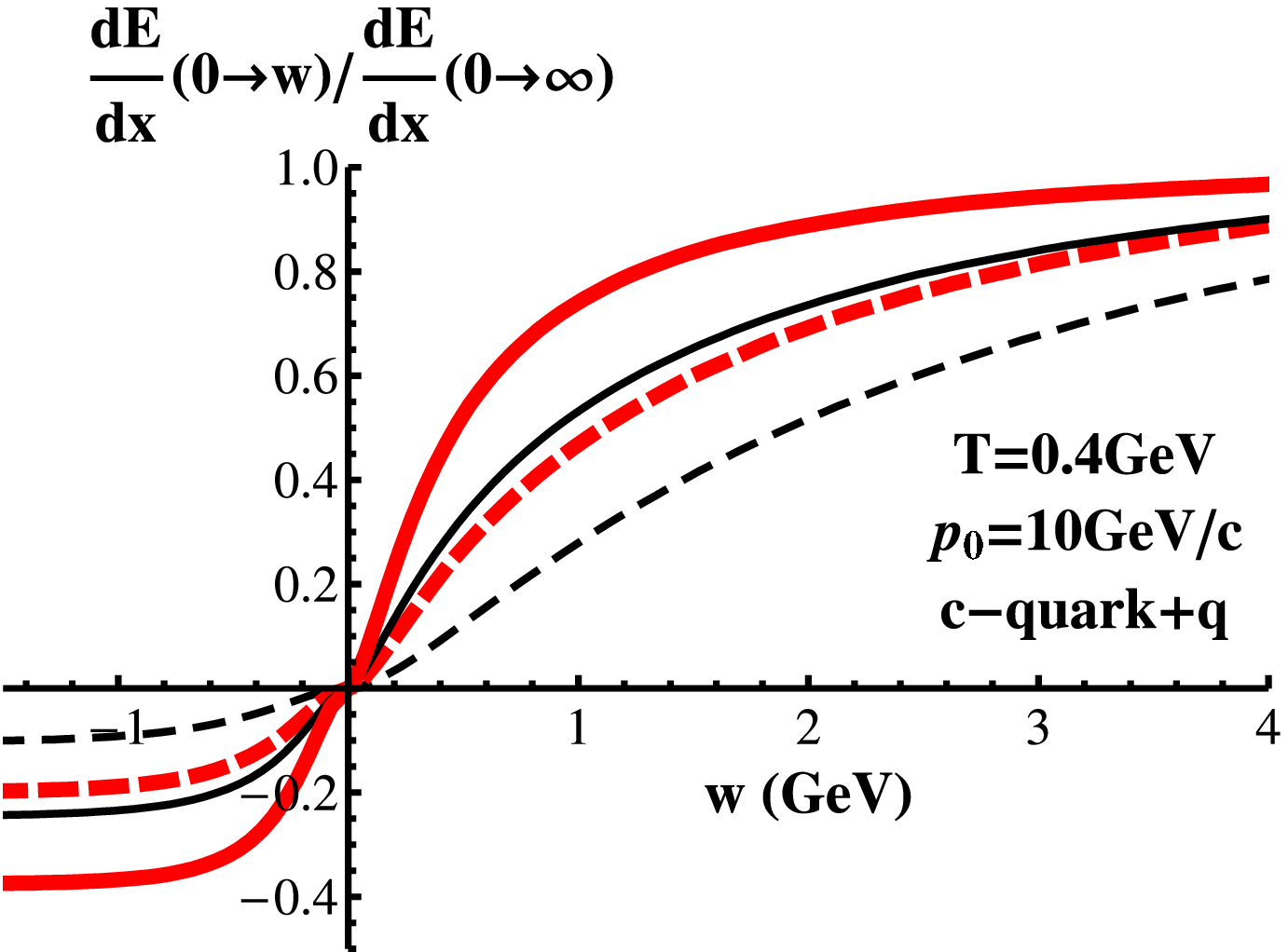,width=0.49\textwidth} \caption{ Differential
energy loss  $ w\,P_q(w) = \frac{dE_q}{dtdw}$ (left) and its
normalized integral (right), both evaluated for a heavy quark with
an initial momentum of $p = 10\,{\rm GeV/c}$ colliding with a quark.
For the different curves see table 1.} \label{pomegb}
\end{figure}
To show which collisions are most important for the total energy
loss of the c-quark we display in fig. \ref{pomegb} (left) (the
absolute value of) $w\,P_q(w,p)$ for cq collisions. (cg collisions
would exhibit a similar behavior). This quantity is directly related
to the differential energy loss: \be
  \frac{dE_q}{dxdw}=
  v^{-1}  P_q(w, p)\,w\,.
\label{eq:doublediffeloss} \ee Collisions with a small
energy transfer become dominant when a running coupling constant
is employed. Fig. \ref{pomegb}, right
shows
\be
 \int^w dw'\frac{dE_q}{dxdw'} / \int^\infty dw' \frac{dE_q}{dxdw'}
\ee and displays that collisions with an energy transfer of $w <1$
GeV contribute 70\% to the total energy transfer in our new approach
whereas in the standard model (B) they contribute 25\% only.

\begin{figure}[htb]%[h]
\epsfig{file=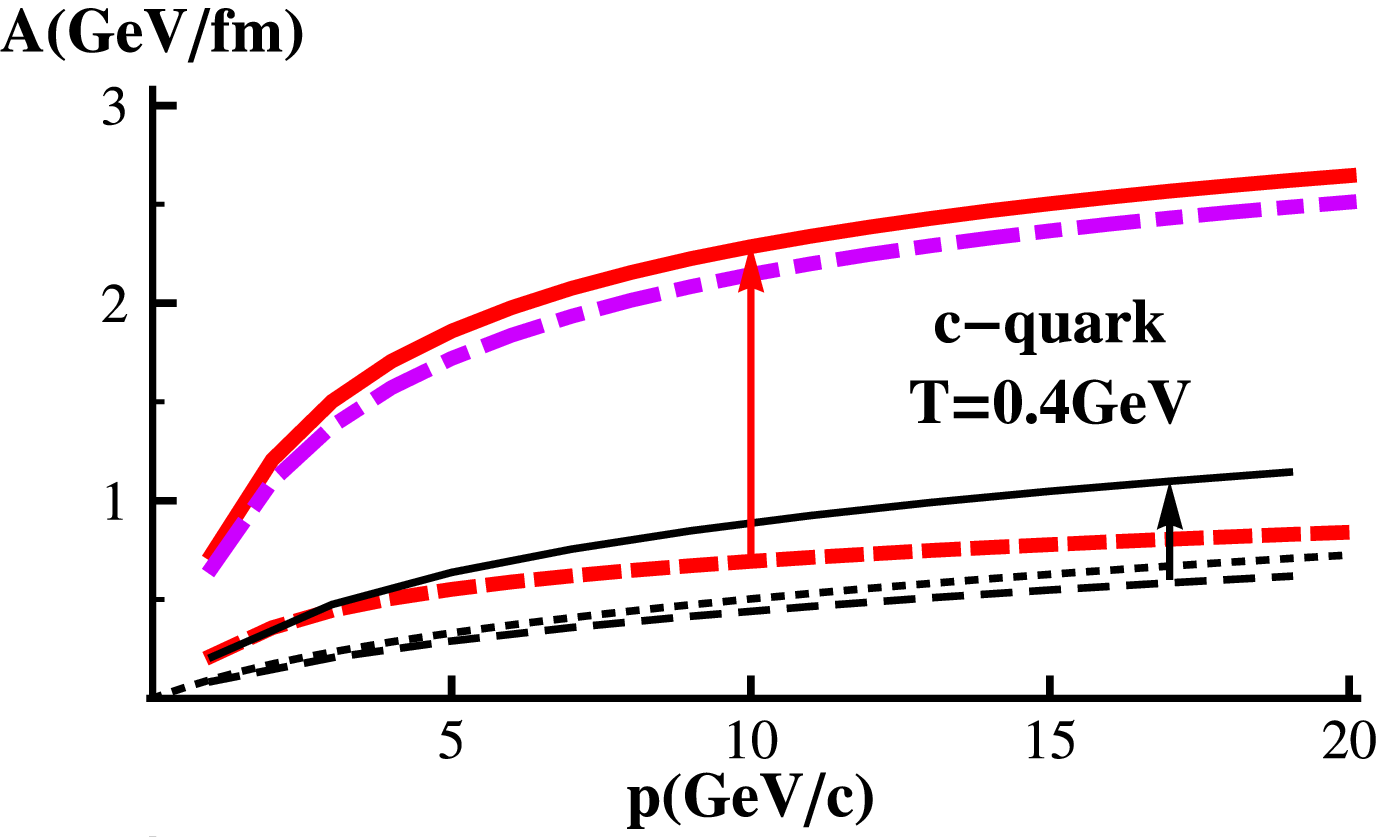,width=0.49\textwidth}
\epsfig{file=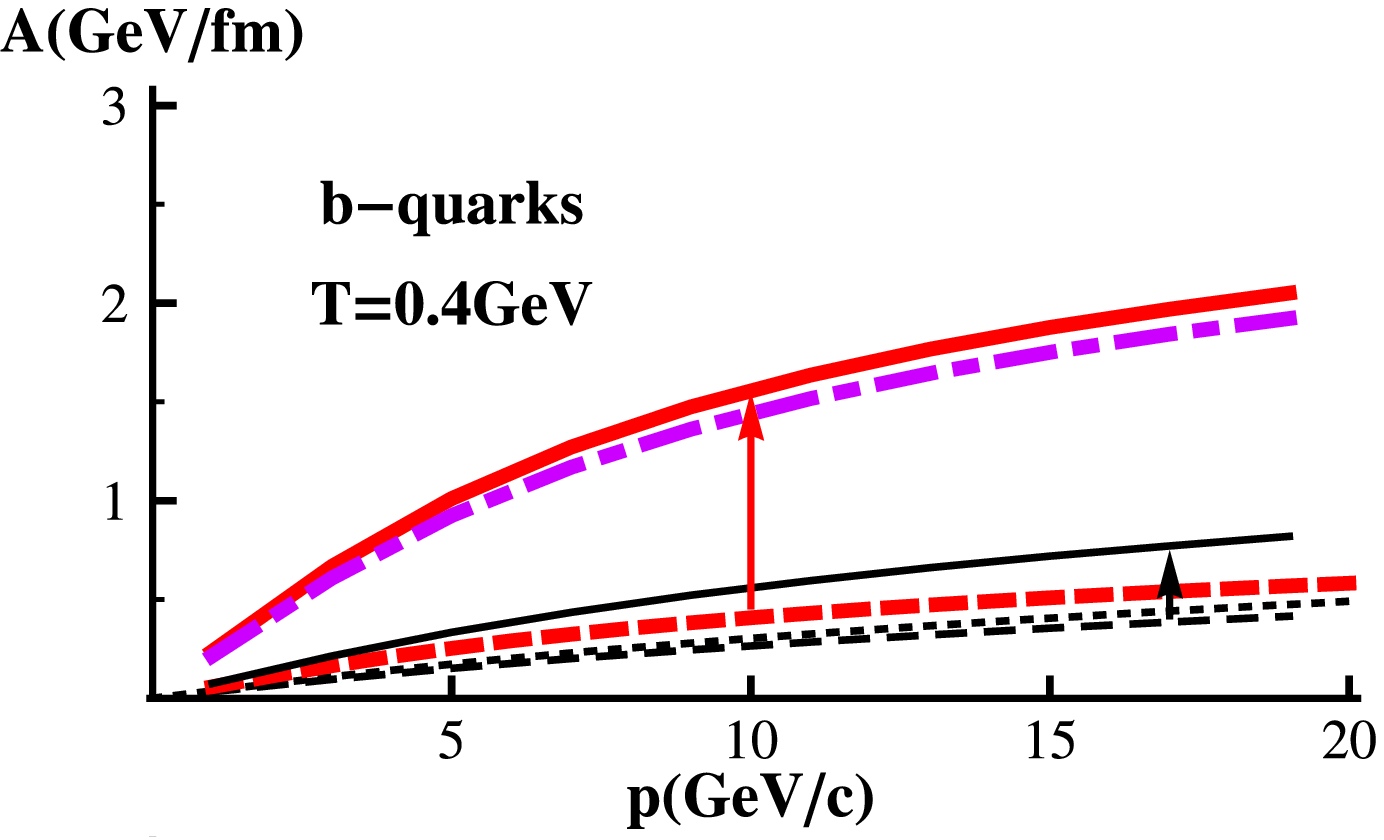,width=0.49\textwidth} \caption{Drag
coefficient A (left for charm quarks, right for bottom quarks) as a
function of the heavy quark momentum $p$. We display A for a
temperature of $T= 400\,{\rm MeV}$ and for different combinations of
coupling constants and infrared regulators as defined in table 1.
For the different curves see table 1.}
% dotted: A, dashed
%thin: B, full thin: C, dashed thick: D, full thick: E and dashed
%dotted: F. The arrows show how the drag coefficients change if one
%replaces $m_D$ by the IR regulator determined by the HTL + semi-hard
%approach.}
\label{asa}
\end{figure}

In order to make our calculation comparable with other
Fokker-Planck calculations we present in fig.
\ref{asa} the drag coefficient A as a function of the Q-quark
momentum $p$ (left for c-quarks and right for b-quarks).
The calculation for the two fixed
coupling constants \as=0.3 and $\alpha_S(2\pi\,T)$ do not yield
different drag coefficients as long as the IR regulator is the same.
Therefore we do not pursue model A. If one changes the IR regulator
from the standard value, $m_D^2$, to that reproducing the HTL energy
loss ($\kappa=0.15$) one observes an increase by a factor of 2. A
running \as $(\alpha_{\rm eff})$ with a standard IR regulator
increases the drag coefficient for low momenta where the small-t
exchanges are more important. If the low t collisions are enhanced
by both, a running \as and a small IR regulator, we see an increase
of the drag coefficient by a factor of $\approx 5$. The drag changes
not substantially if the IR regulator is calculated with a running
coupling constant -- model F -- as compared with model E and we
therefore discard model F from further calculations. If \as remains
fixed the drag coefficient remains moderate for all infrared regulators,
as it does for a running \as and the Debye mass as infrared regulator.
B-quarks show a similar behavior but their drag coefficient is - due
to their higher mass - around 30-40\% smaller than that of the
c-quarks. For a given plasma-lifetime evolution, we thus expect a
smaller energy loss of b quarks, but it is far from  being
negligible, especially in the most realistic models E and F.

\begin{figure}[htb]%[h]
\epsfig{file=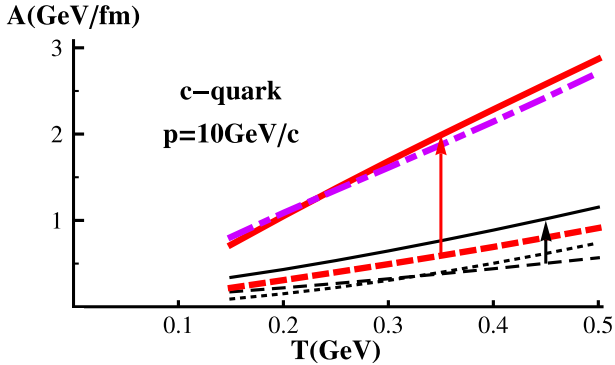,width=0.49\textwidth} \caption{Temperature
dependence of the drag coefficient A for c-quark with momentum$
p=10\ {\rm GeV/c}$. For the different curves see table 1.
%The symbols are the same as in fig. \ref{asa}.
The arrows
show how the drag coefficients change if one replaces $m_D$ by the
IR regulator determined by the HTL + semi-hard approach.}
\label{asb}
\end{figure}

The drag coefficient depends strongly on the temperature. In
fig.\ref{asb} we display that of a c-quark with a momentum of 10
GeV/c. As expected in our model, a hot plasma is much more effective
to quench a fast quark than a cold one.

\subsection{Nucleus-Nucleus Collisions}
\label{sect_indiv_coll} After having discussed single Qq and Qg
collisions we investigate now the consequences of our approach for
heavy quark observables in ultrarelativistic heavy ion collisions.
To study the time evolution of the heavy quark in a plasma, usually
a Fokker-Planck equation has been used. This approach has several
shortcomings: a) The drag and diffusion coefficients, calculated by
eq.\ref{eq:meanv}, do fulfill the Einstein relation only in leading
logarithmic order E/T \cite{Moore:2004tg}. This is not sufficient to
assure the thermalization of the heavy quark \cite{Walton:1999dy}.
Either one has to impose the Einstein relation or the asymptotic
heavy quark distribution is a Tsallis function and not a Boltzmann
distribution. b) Being a small scattering angle approximation (or,
in other words, containing the leading order term of $T/E_Q$ only)
the approach brakes down if the momenta of the q(g) and of the Q are
of the same order, i.e. in the region where $v_2$ becomes large. c)
Even for large energies $E_Q$ first and second moment only
(eq.\ref{eq:meanv}) are not a good approximation to the energy loss.
It can be seen in fig.\ref{cross} (right) that hard transfers are
not excluded in the gluonic channel, due to the QCD-equivalent of
the Compton effect.

Therefore, for the calculation presented here, we use a Boltzmann
equation approach as in ref.\cite{Molnar:2004ph} in a test particle
version. In coordinate space the initial distribution of the heavy
quarks is given by a Glauber calculation. For the momentum space
distribution as well as for  the relative contribution of charmed
and bottom quarks we use the pQCD results of \cite{Cacciari:2005rk}.
In the E866 experiment at Fermi Lab \cite{E866} it has been observed
that in pA collisions J/psi mesons have a larger transverse momentum
as compared to pp collisions. This effect, called Cronin effect, can
be parameterized as an increase of $<p_T^2>$ by $\delta_0\approx
(0.2\ {\rm GeV})^2$ per collision of the incident nucleon with one
of the target nucleons. For most of the calculations we then
convolute the initial transverse-momentum distribution of the heavy
quark \cite{Cacciari:2005rk} with a Gaussian of r.m.s $\sqrt{n_{\rm
coll}(\vec{r}_{\perp})\,\delta_0}$. In this parametrization $n_{\rm
coll}$ is taken as the mean number of soft collisions which the
incoming nucleons has suffered prior to the formation of the
$Q\bar{Q}$ pair at transverse position $\vec{r}_{\perp}$. Future
studies of D/B meson production at RHIC may allow to improve on this
choice.

In our approach we then follow the trajectories of the individual heavy
quarks in the expanding plasma, described by the hydrodynamical
model of Kolb and Heinz \cite{heko,Gossiaux:2006yu}. We parameterize
the temperature $T(r,t)$ and the velocity $u_\mu(r,t)$ field of this
model and use this parametrization in a finite time step method to
calculate the collision rate $\Gamma$ ( eq. \ref{eq:meanv} with $X=
1$) for $Q+g \to Q+g$ and $Q+q\to Q+q$ reactions
(\cite{Svetitsky:1987gq,Combridge:1978kx}) and for the different
parameterizations of the cross section. For a given interval of the
(Bjorken) time $\Delta \tau$, we then generate the number of
collisions according to a Poisson distribution of average $\Gamma
\Delta \tau$ and perform these collisions individually.
When a collision takes place we determine the final momentum of the
heavy quark by taking randomly a scattering angle with a
distribution given by the cross section at a given temperature. In
this method no small angle approximations are necessary and we
arrive by definition at a thermal distribution if we place the
Q-quark in infinite matter at a given temperature.

As the time-point of the hadronization of the plasma is not well
determined, we explore here two options: a) Hadronization of heavy
quarks into D(B) mesons when the expanding system enters the mixed
phase and b) at the end of the mixed phase. In the latter option
more collisions are possible and we expect therefore a larger
quenching of heavy quarks. Also for the hadronization we apply
two approaches which give slightly different meson momentum distributions:
a) either we apply exclusively the fragmentation mechanism
as in p-p \cite{Cacciari:2005rk} or b) we apply the fragmentation
mechanism for high momentum quarks only whereas at low momentum heavy
mesons are formed by coalescence. For this purpose we define the probability
distribution $g$ that
a heavy meson of momentum $\vec{P}$ is formed by coalescence of a
heavy quark with momentum $\vec{p}_Q$ with a light quark as \be
g(\vec{P},\vec{p}_Q)= \beta \int d^3q \,
n(q,T)f(\vec{q}-\vec{p}_Q)\delta(\vec{P}-\vec{p}_Q -\vec{q})\,, \ee
where  $n(q,T)$ is the thermal momentum distribution of the light quarks
at the moment of hadronization and f is the probability
density that the heavy quark with a momentum $\vec{p}_Q$ forms a
heavy meson with a light quark of momentum $\vec{q}$. In the
calculation we evaluate $g$ in the fluid rest frame and take f as a
boosted Gaussian. $\beta$ is chosen such that $g$ is normalized to
unity for $\vec{p}_Q = 0$.  Finally the heavy meson undergoes a weak
decay and creates the single electrons which are observed in the
detector.

\begin{figure}[htb]%[h]
\epsfig{file=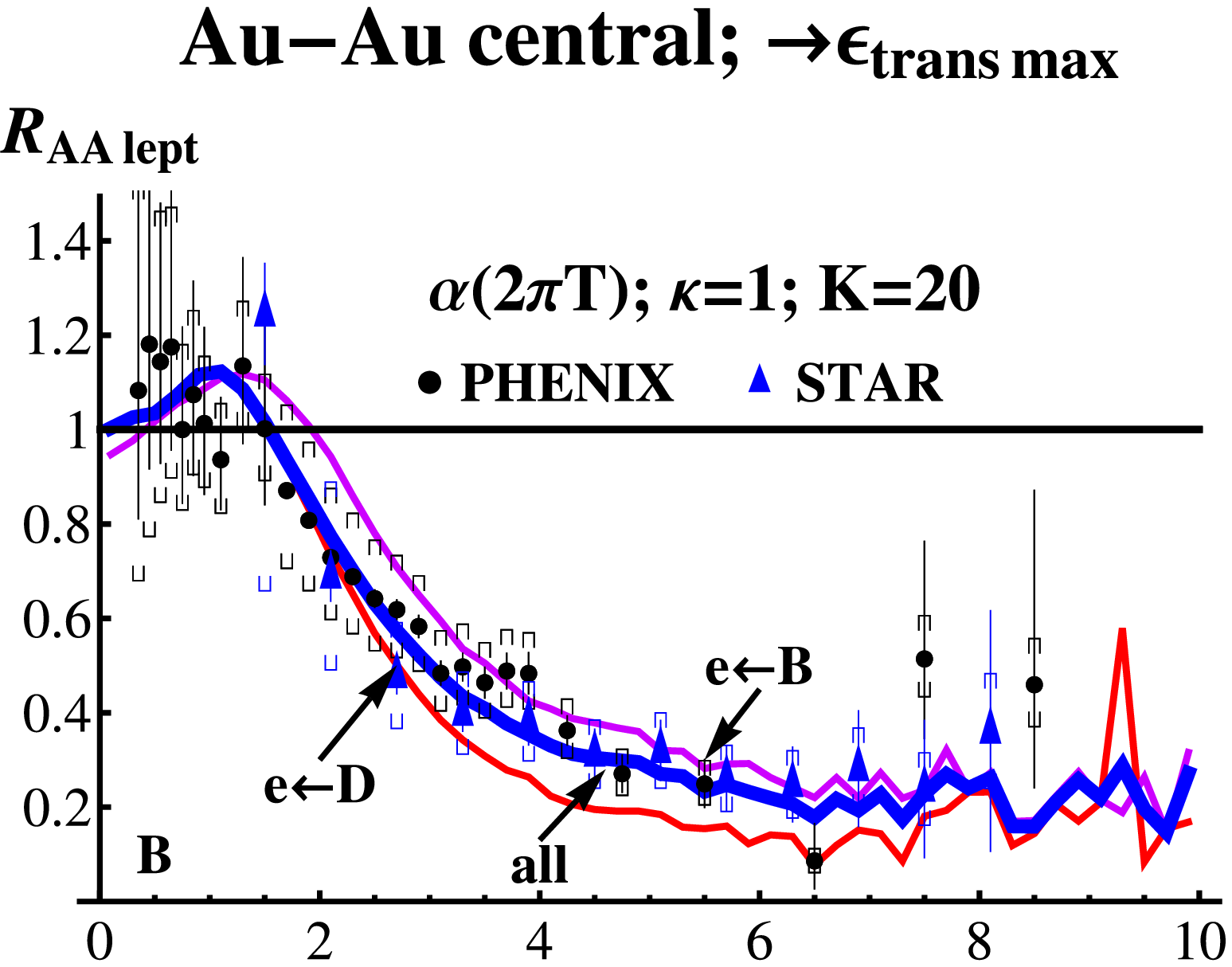,width=0.4\textwidth}
\epsfig{file=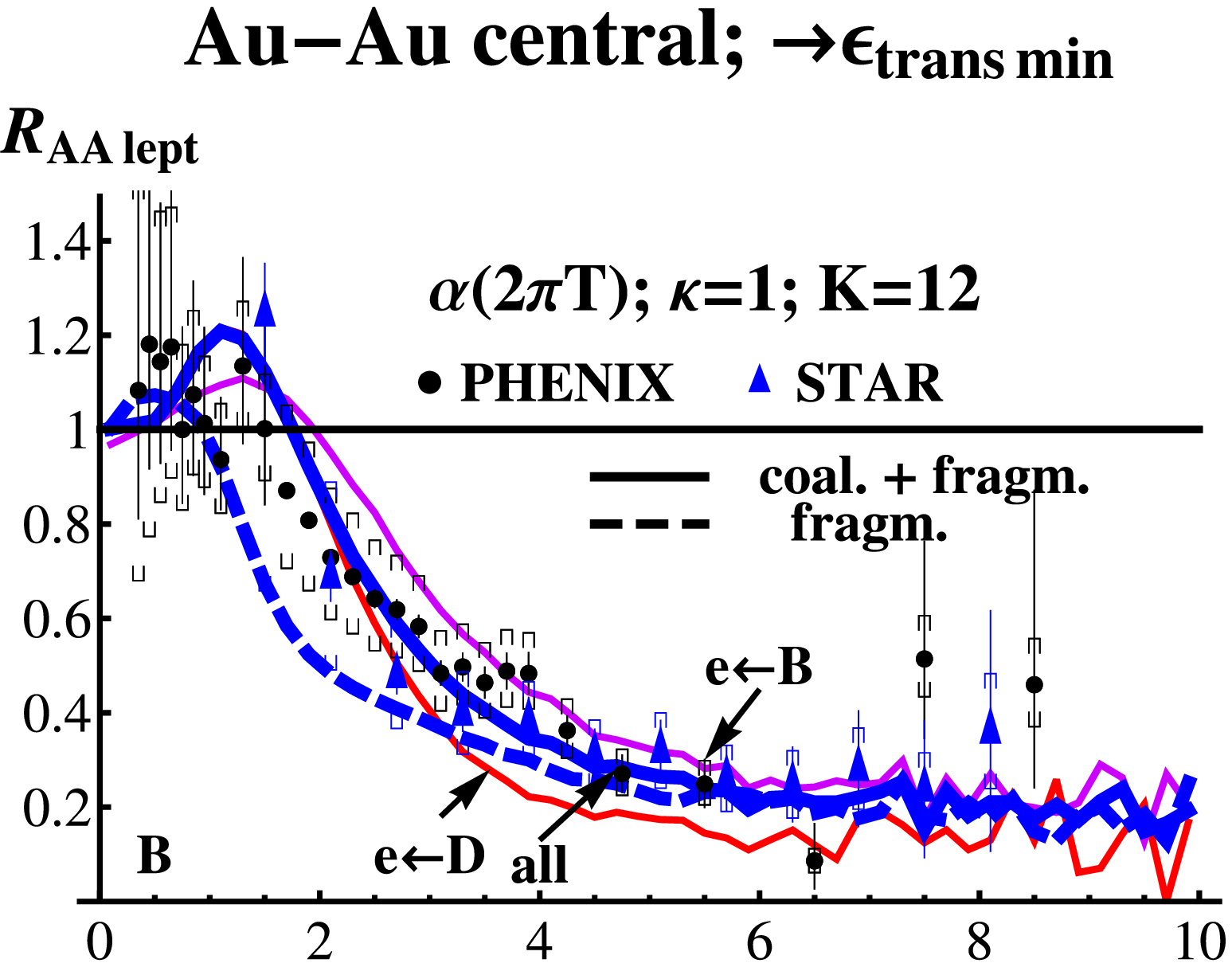,width=0.4\textwidth}
\epsfig{file=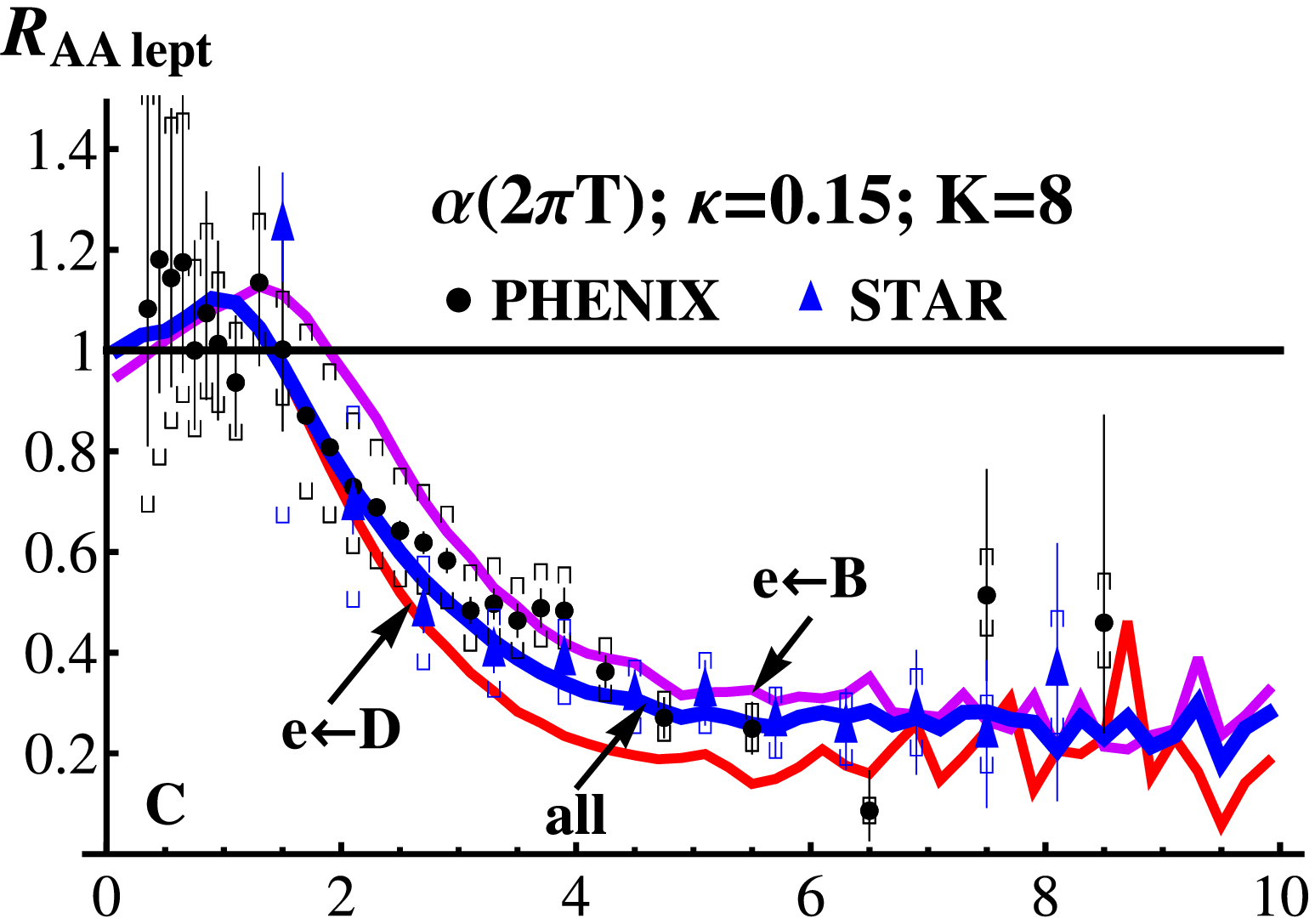,width=0.4\textwidth}
\epsfig{file=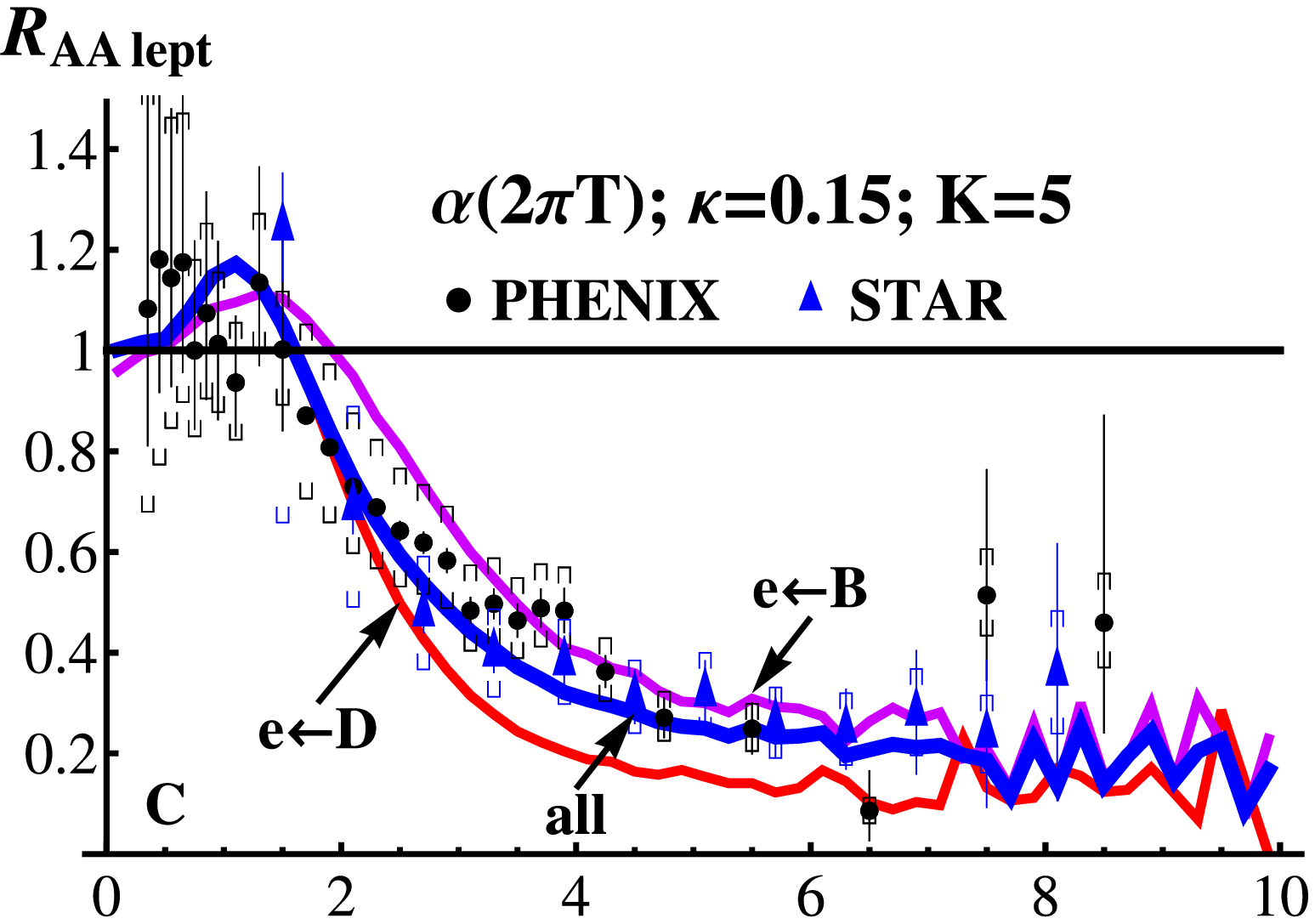,width=0.4\textwidth}
\epsfig{file=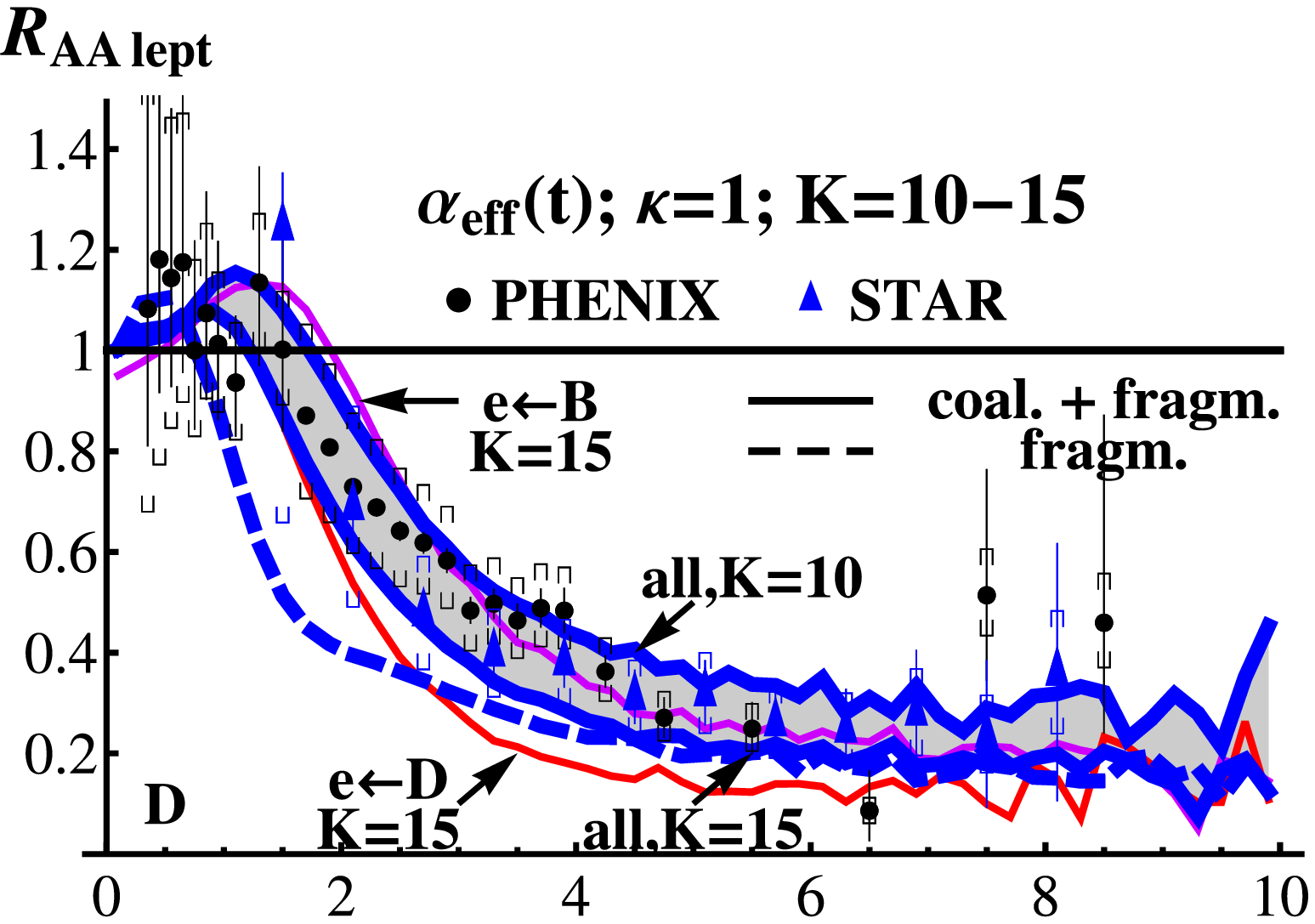,width=0.4\textwidth}
\epsfig{file=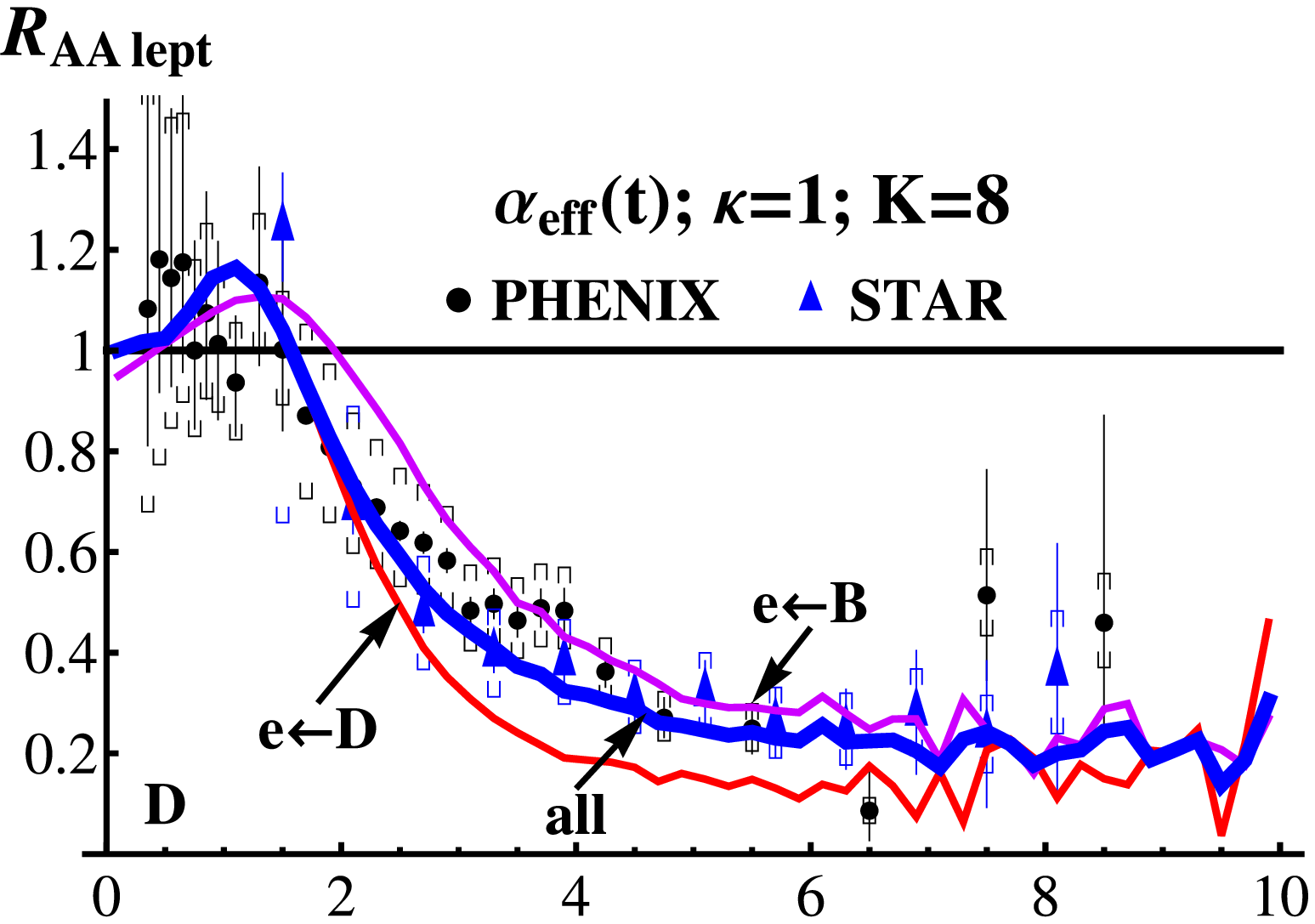,width=0.4\textwidth}
\epsfig{file=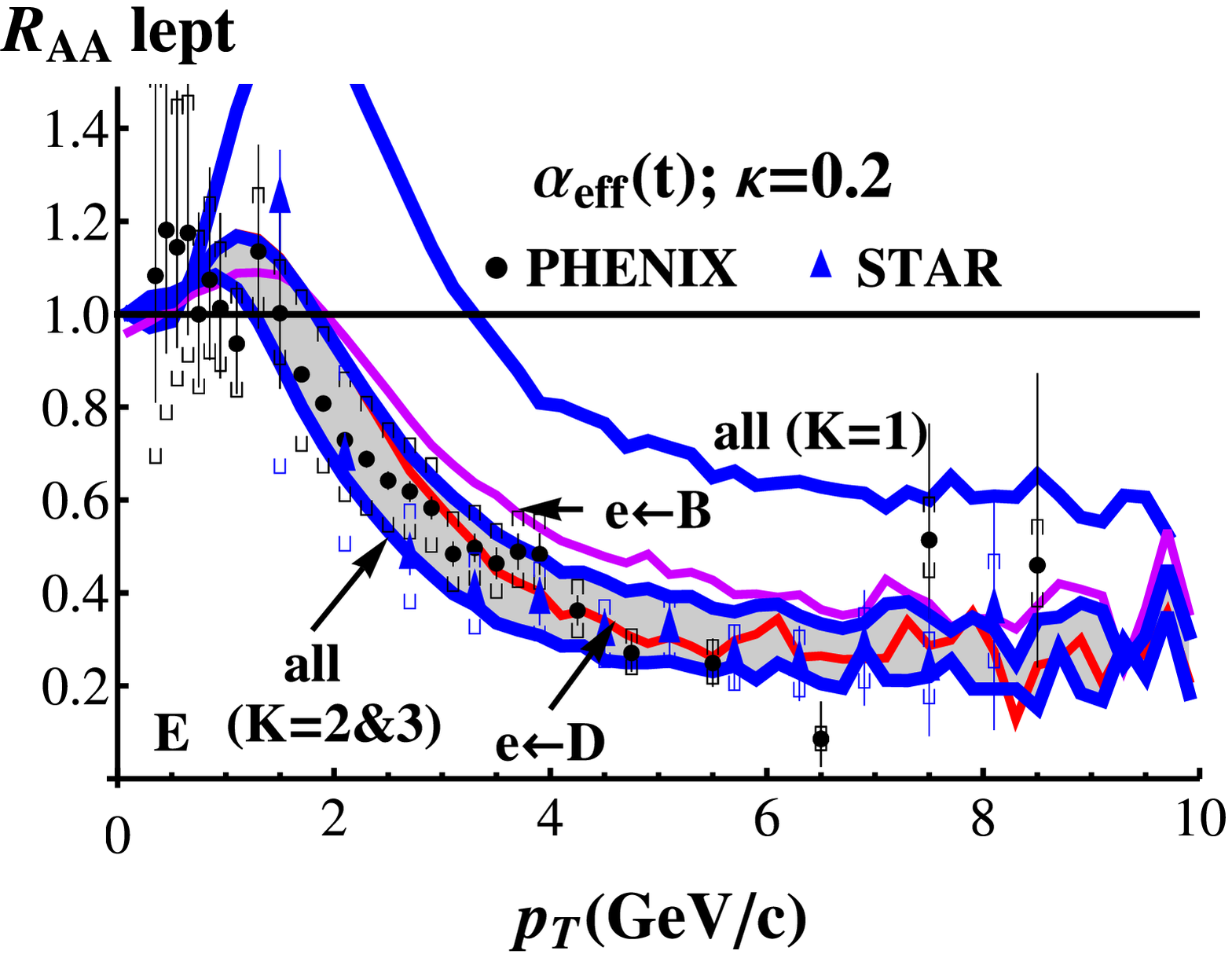,width=0.4\textwidth}
\epsfig{file=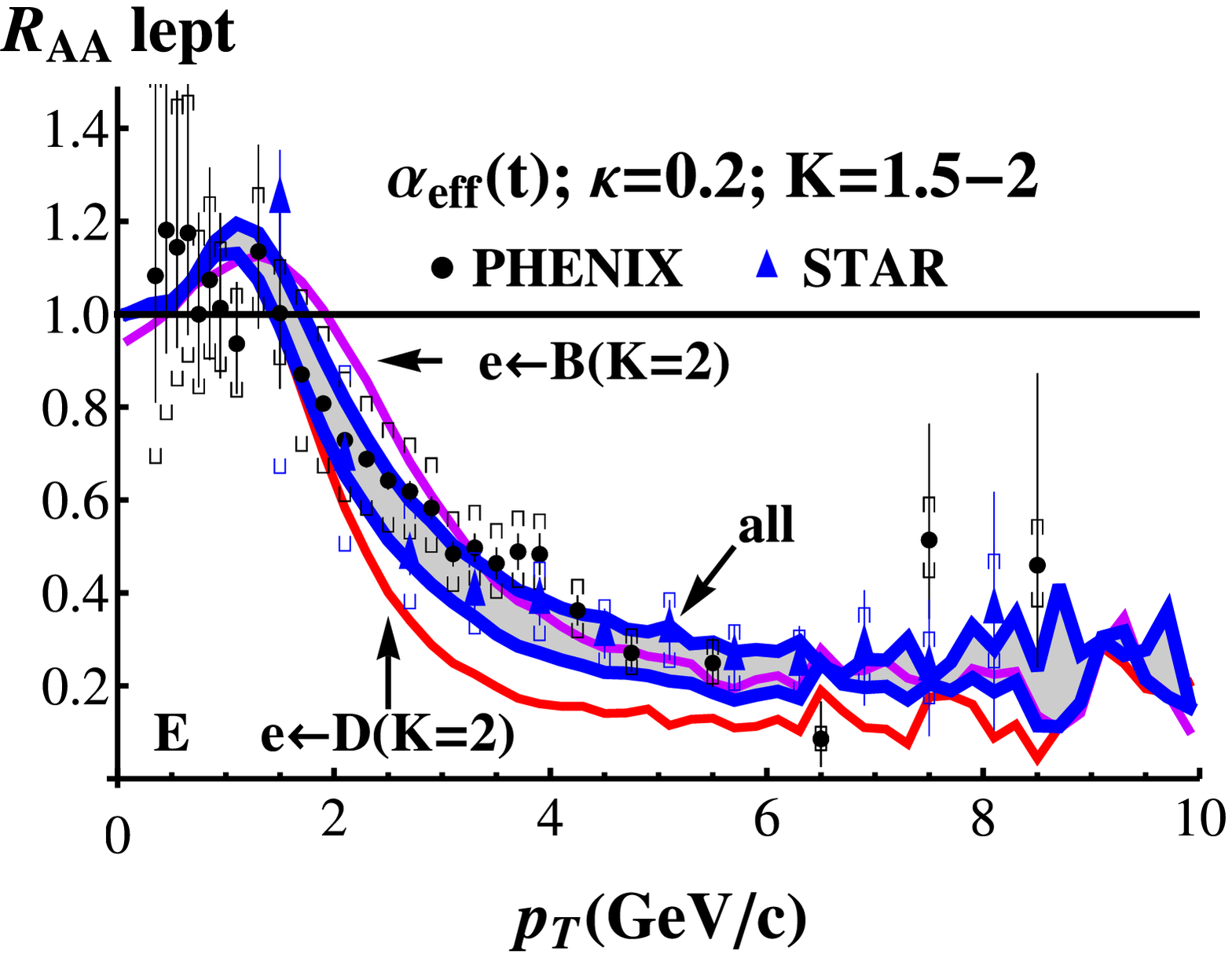,width=0.4\textwidth} \caption{Comparison of
the experimental and theoretical results for central Au+Au
collisions. We display $R_{AA}$ of single non-photonic $e^-$ as a
function of the heavy quark momentum $p_T$. The purple line shows
$R_{AA}$ for $e^-$ from B-meson and the red line that of D-meson
decay for the K values indicated in the figure. The blue line is the
sum of both. On the left hand side we assumed hadronization at the
beginning of the mixed phase on the right hand side at the end of
the mixed phase. From top to bottom we display the results for the
parameterizations B-E (see table 1).} \label{RAAa}
\end{figure}

The results for $R_{AA}$ in central Au+Au collisions are compared
to the experimental data in  fig. \ref{RAAa}. From
top to bottom we show the results for the approaches B-E  of table
1. On the left hand side we present the results for an hadronization
at the beginning of the mixed phase, on the right hand side that for
an hadronization at the end of the mixed phase. We observe that the
additional interactions in the mixed phase reduce the artificial K
factor, shown in the figure, with which the pQCD cross section has
to be multiplied to describe the data. For some of the curves we
present the results for 2 different values of K, in others we show
the influence of the different approaches for fragmentation.
``frag'' means that heavy mesons are exclusively created by
fragmentation, ``coal. + frag.'' means that they are rather produced
by coalescence at low momentum. It is evident that the different
hadronization scenarios have little influence on the K factor which
is necessary to describe the data. For a constant coupling constant
and the Debye mass as IR regulator (model B) one has to employ
K-factors of the order of 10-12. A smaller IR regulator (model C) or
a running coupling constant (model D) reduce this K factor to values
of 5-10, still much too large in order to render the calculation
understandable. Only the combination of both, model E, brings the K
factor close to an acceptable value of 1-2, leaving nevertheless
still room for radiative energy loss.

\begin{figure}[htb]%[h]
\epsfig{file=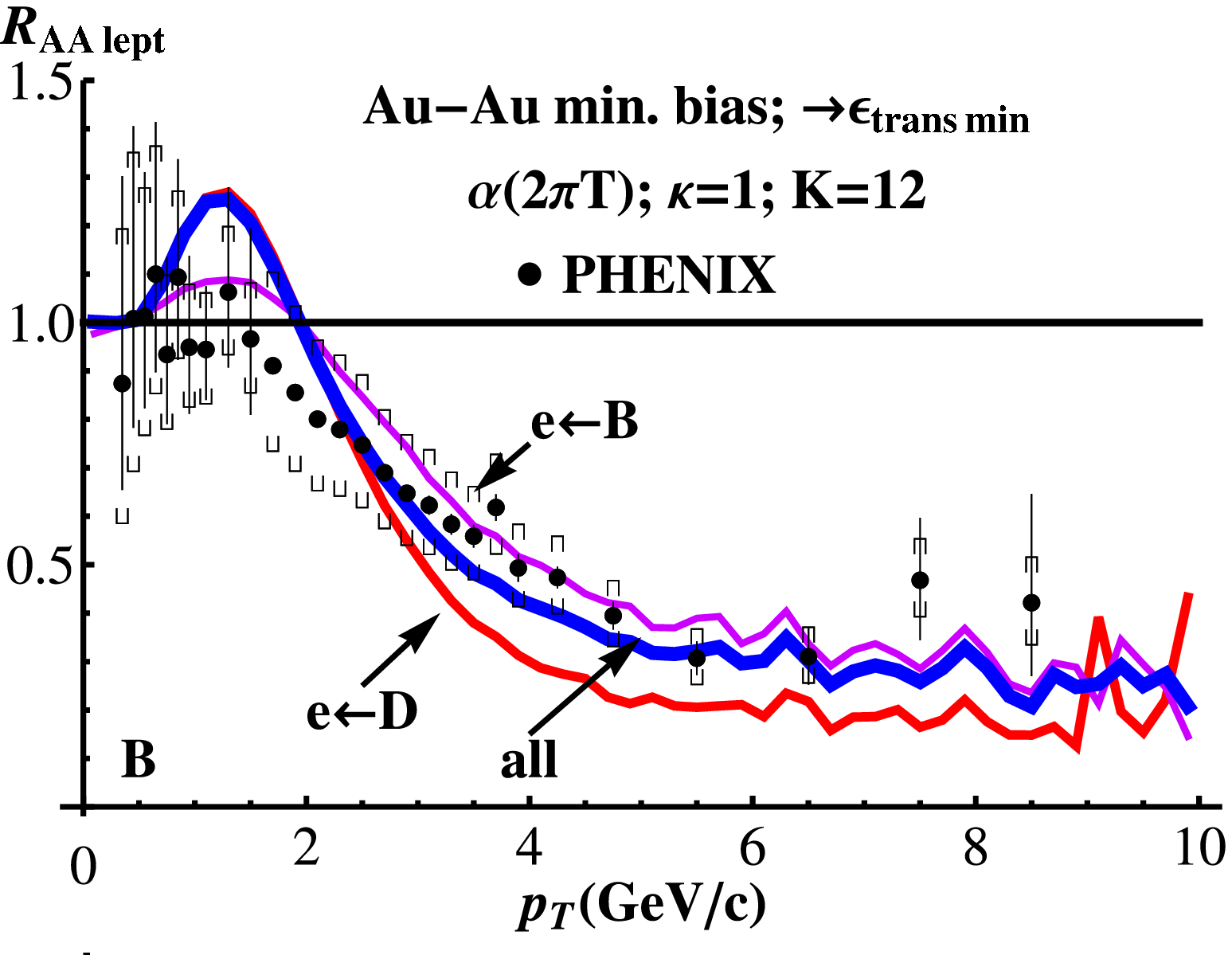,width=0.49\textwidth}
\epsfig{file=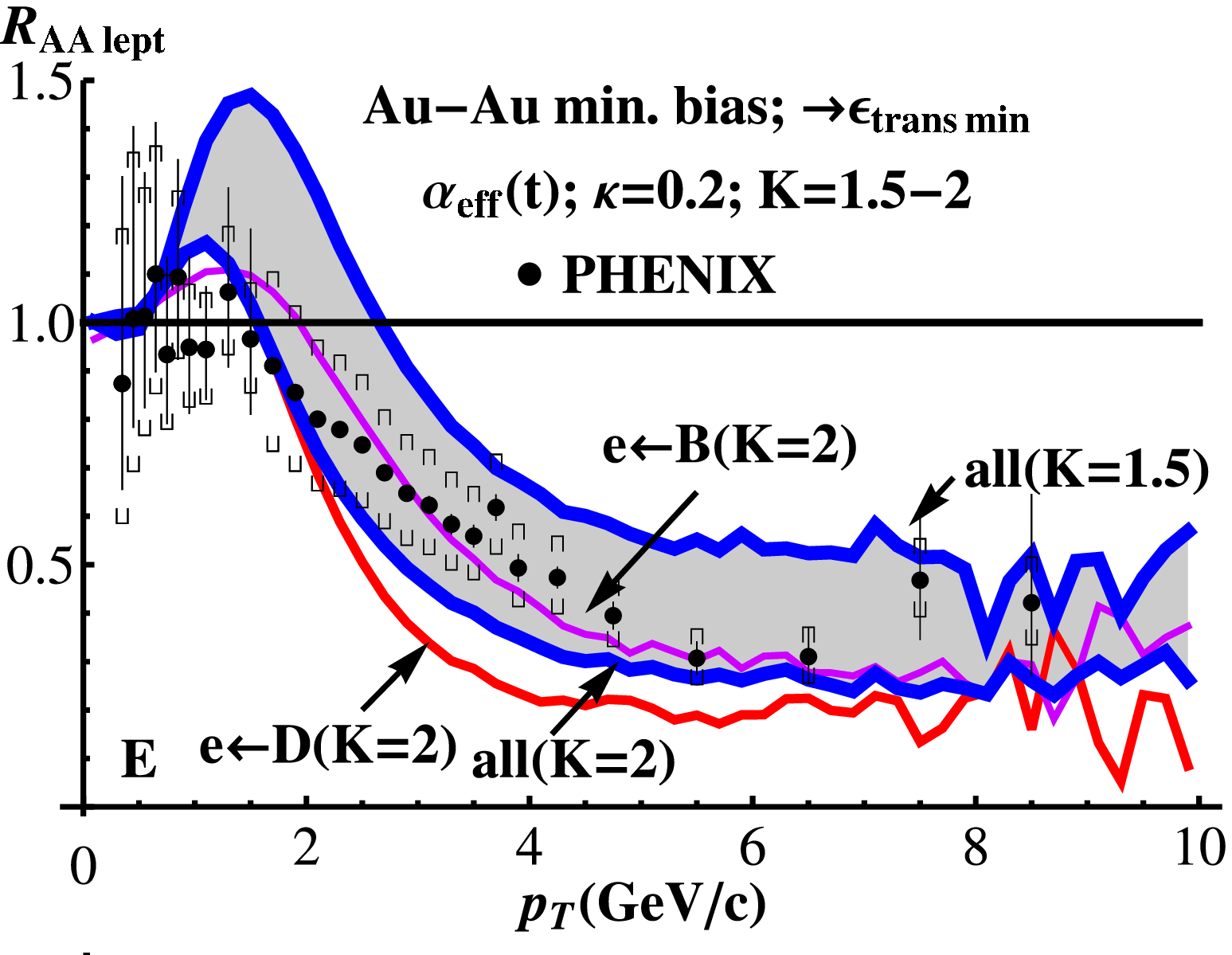,width=0.49\textwidth} \caption{Comparison of
the experimental and theoretical results for minimum bias Au+Au
collisions. We display $R_{AA}$ of single non-photonic $e^-$ as a
function of the heavy quark momentum $p_T$. The red line shows the
$e^-$ for D-meson decay, the purple line those for B-meson decay and
the blue thick line the sum of both. Hadronization is assumed to
take place at the end of the mixed phase. On the left hand side we
display the results of model B, on the right hand side that of model
E (see table 1). The applied K factors are given in the figure, the
Cronin effect is taken into account.} \label{RAAb}
\end{figure}

A very similar observation can be made for the minimum bias
calculations which are compared  with the experimental data in fig.
\ref{RAAb}. On the left hand side we display the results for model
B, on the right hand side for model E. For a fixed coupling constant
and the Debye mass as the infrared regulator we need, as for central
collisions, a K factor of around 12, whereas for the model E the K
factor is reduced to 1.5-2. Thus for central and minimum bias
calculations the same K-factors have to be employed, a minimal
requirement for the validity of this reaction scenario.

\begin{figure}[htb]%[h]
\epsfig{file=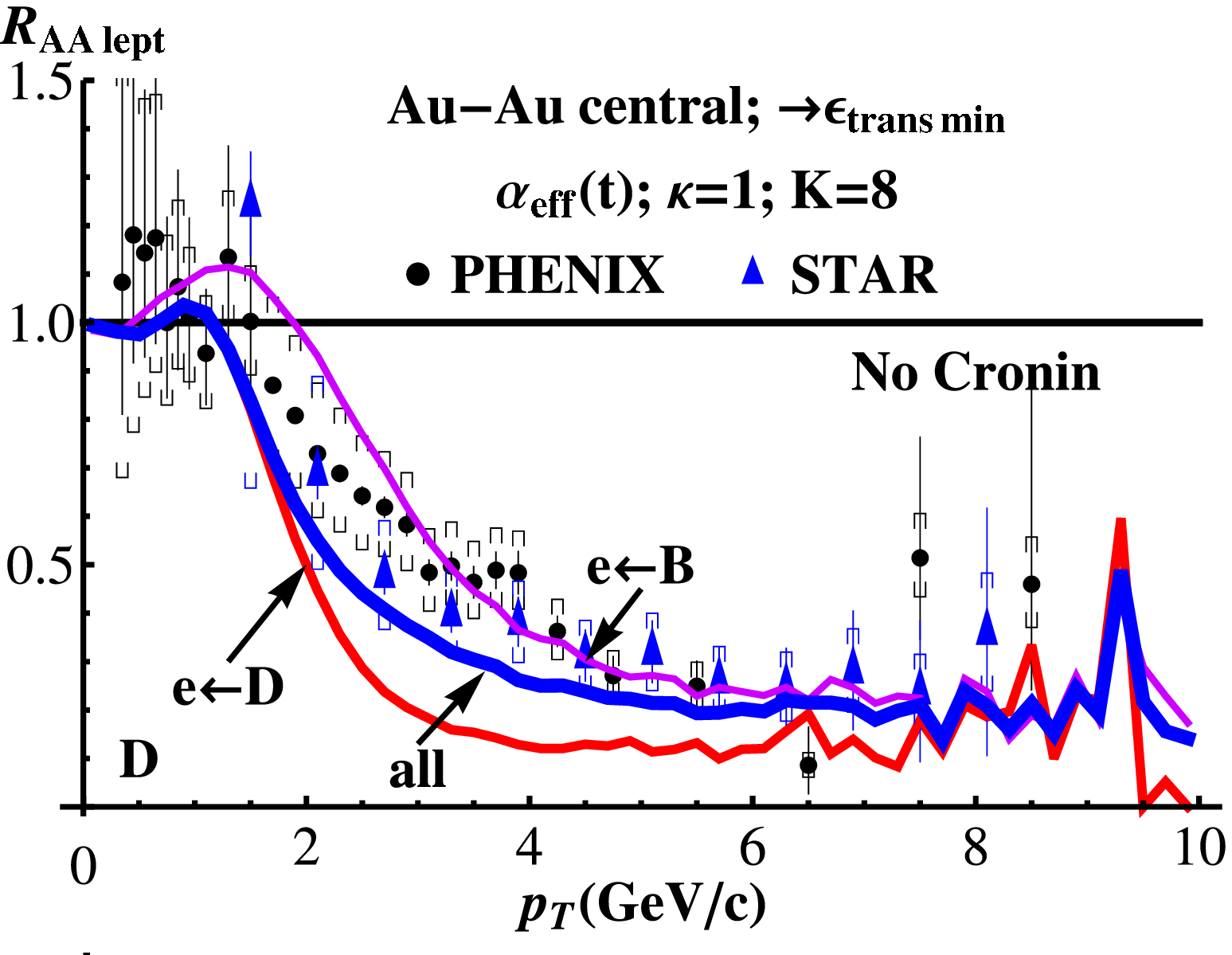,width=0.49\textwidth}
\epsfig{file=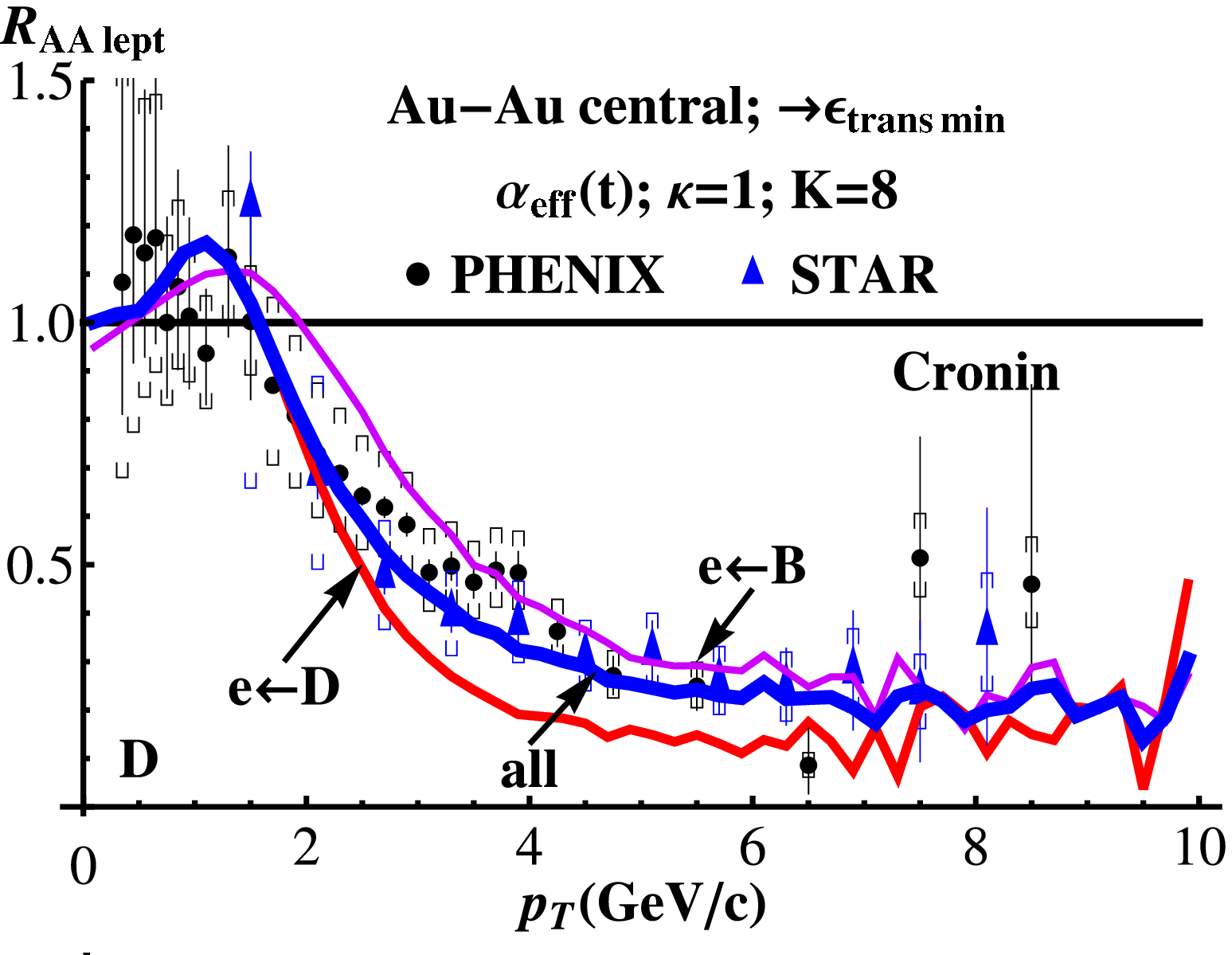,width=0.49\textwidth} \caption{Influence of
the Cronin effect on the $R_{AA}$ of single non-photonic $e^-$ as a
function of the heavy quark momentum $p_T$. The red line shows the
$e^-$ from D-meson decay, the purple line those from B-meson decay
and the blue thick line the sum of both. We assumed hadronization at
the end of the mixed phase. On the left hand side we display the
results without, on the right hand side with the Cronin effect, both
for the model E (see table 1).} \label{RAAc}
\end{figure}

The  Cronin effect changes the  $R_{AA}$ value only for momenta
between 1 and 3 GeV, as can be seen in fig. \ref{RAAc}. It is
therefore without any importance for the understanding of the
$R_{AA}$ values at large $p_T$ but brings $R_{AA}$ much closer to
the data in the $p_T$ range where the $v_2$ values are large.
\begin{figure}[htb]%[h]
\epsfig{file=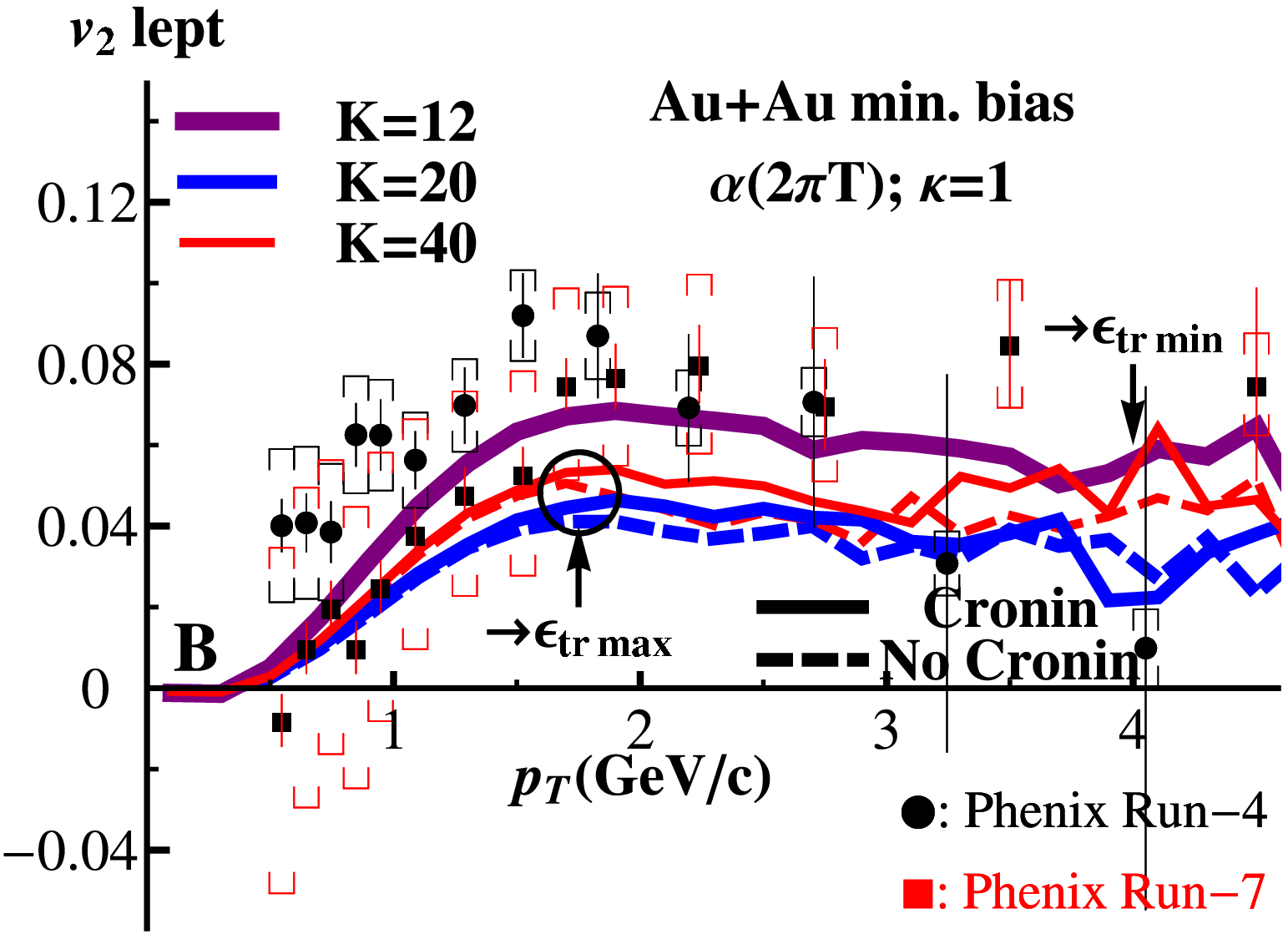,width=0.49\textwidth}
\epsfig{file=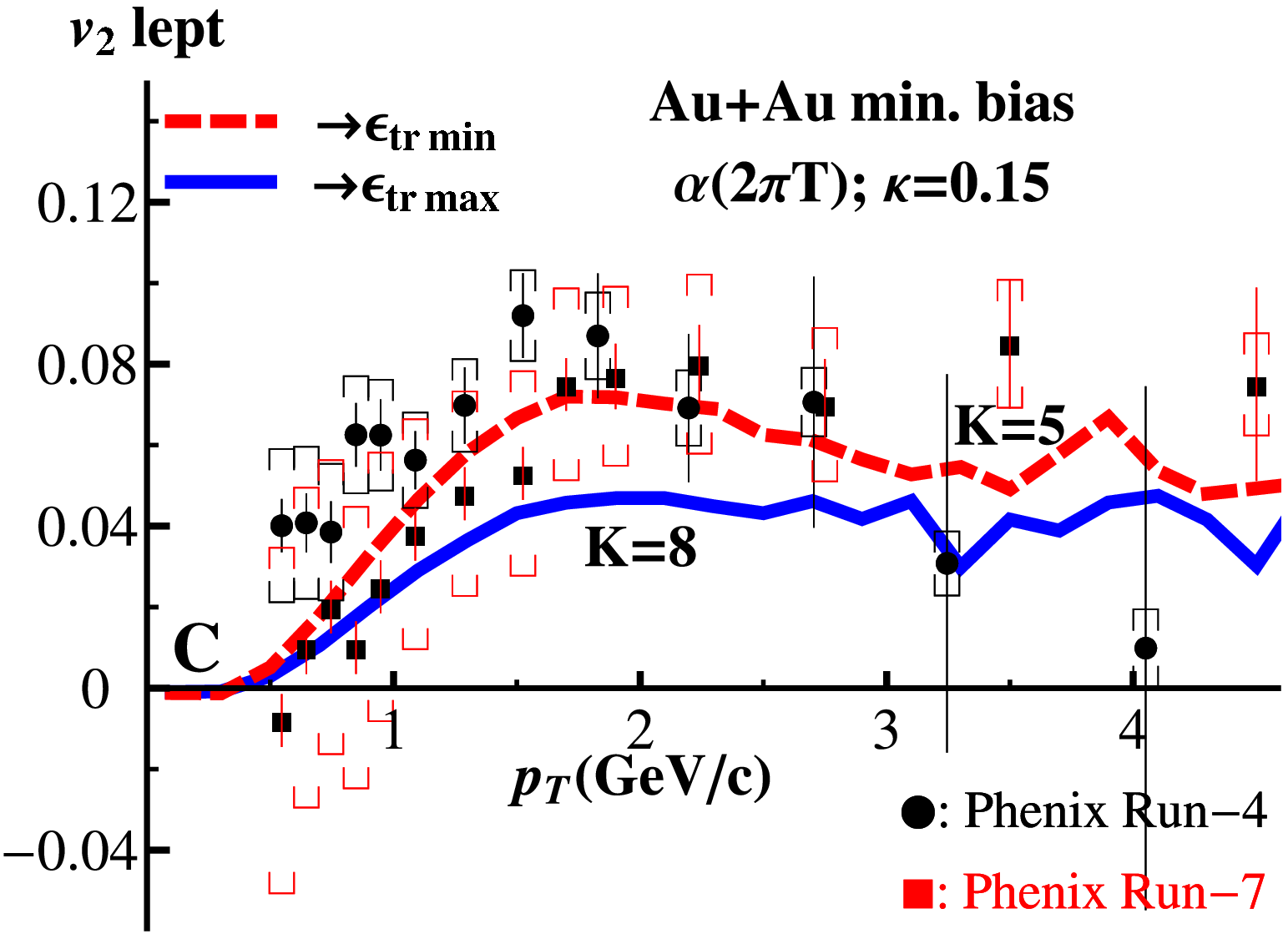,width=0.49\textwidth}
\caption{Dependence of the $v_2$ of single non-photonic  $e^-$ as a
function of the heavy quark momentum $p_T$ on the Cronin effect and
on the K factor (left) as well as on the freeze out density and on
the infrared regular (right). All calculations are done with
$\alpha_S(2\pi T)$.} \label{v2a}
\end{figure}

\begin{figure}[htb]
\epsfig{file=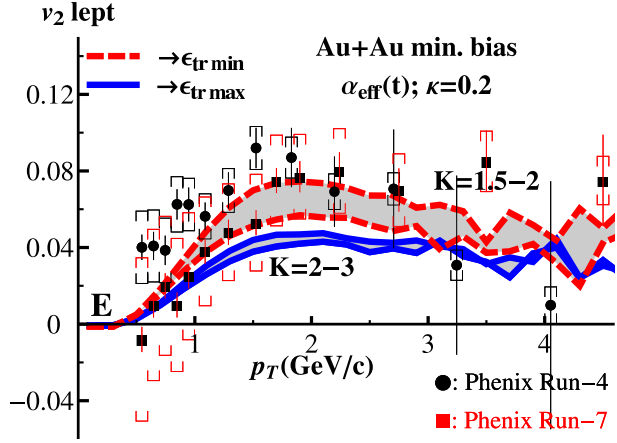,width=0.49\textwidth} \caption{$v_2$
of single non-photonic  $e^-$ as a function of the heavy quark
momentum $p_T$ for different freeze out energies using a running
coupling constant and a small infrared regulator (model E).}
\label{v2b}
\end{figure}

We come now to the discussion of $v_2$. To our knowledge, the
present theories based on pQCD have not succeeded to describe {\it
simultaneously} the experimental $R_{AA}$ and $v_2$ results. As
shown in \ref{v2a}, left, for model B neither the Cronin effect nor
an augmentation of the K factor beyond the value needed to describe
$R_{AA}$ increases $v_2$ considerably. What helps is a larger
interaction time, i.e. a late freeze out. This is shown in
fig.\ref{v2a}, right, where we compare the $v_2$ values for a
hadronization at the beginning and at the end of the mixed phase.
Using a fixed coupling constant the K-factors remain, however,
large. If one combines a running \as with a HTL + semi-hard infrared
regulator one can reproduce $v_2(p_T)$ using a K-factor slightly
larger than $2$ and assuming a late freeze out, as can be seen in
fig. \ref{v2b}.

\begin{figure}[htb]
\epsfig{file=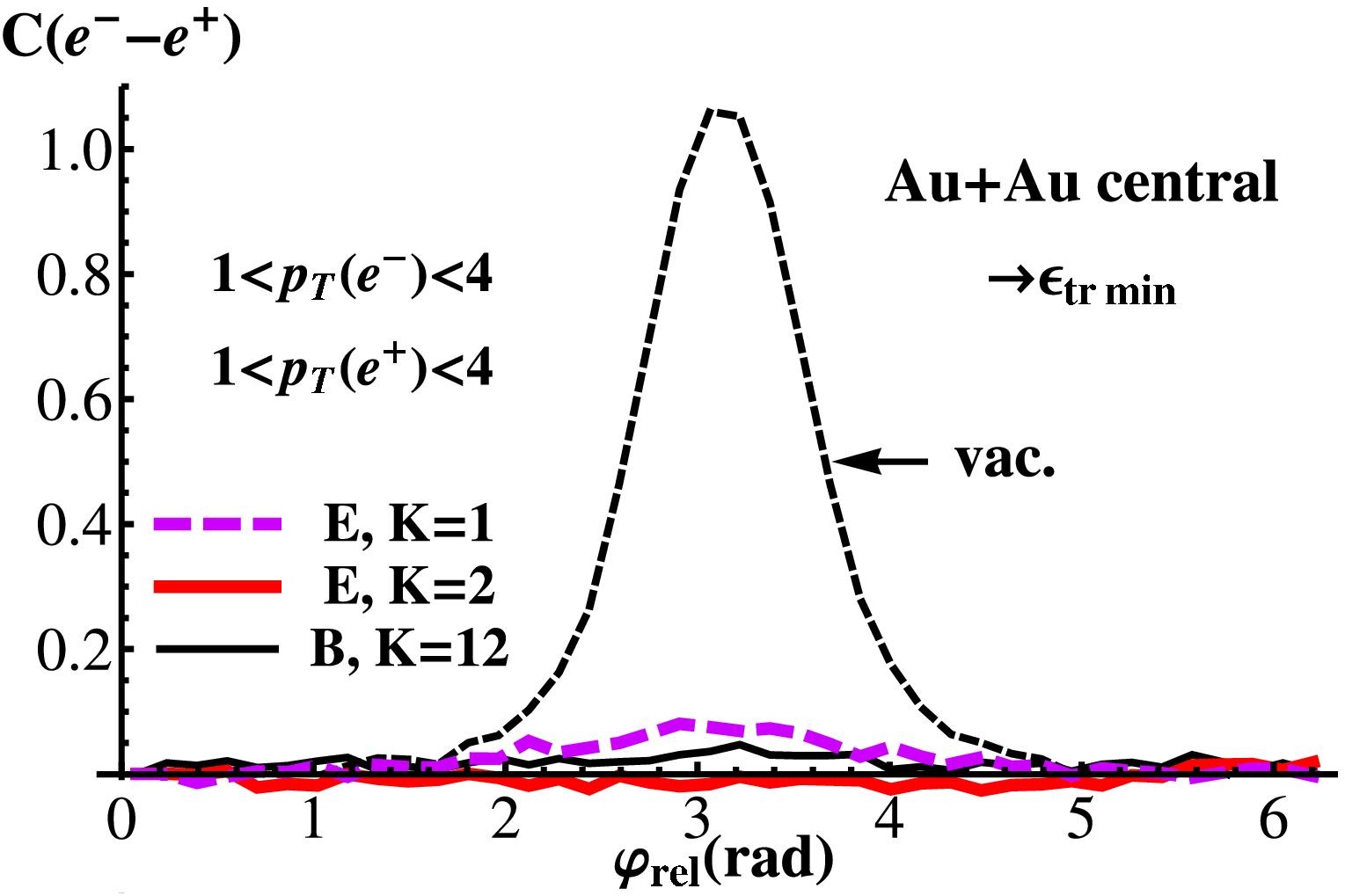,width=0.49\textwidth}
\epsfig{file=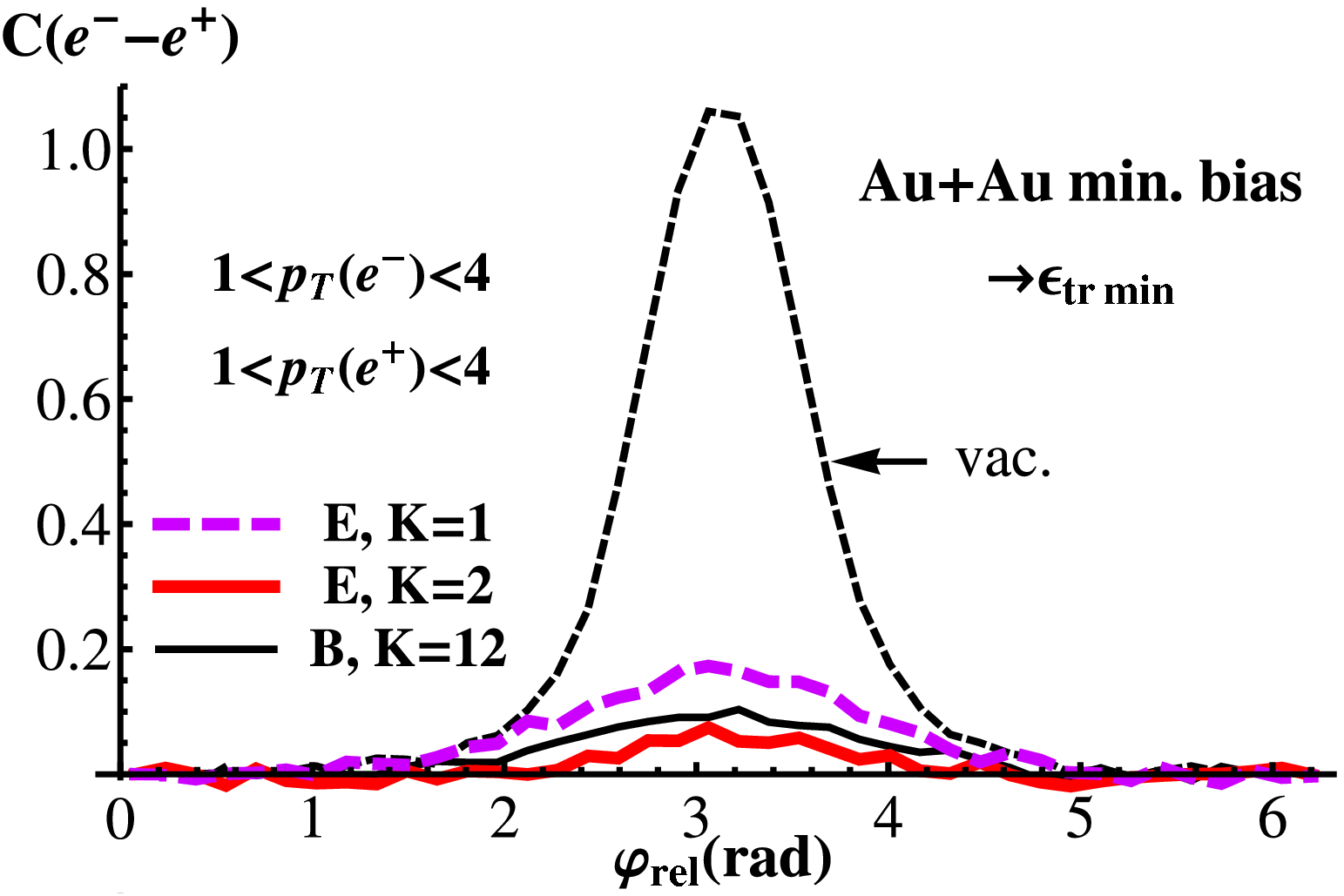,width=0.49\textwidth}
\caption{Azimuthal correlation of $e^+-e^-$ non-photonic pairs as a
function of the relative angle for model E and for both,
central(left) and minimum bias (right), collisions; $Q$ and $\bar{Q}$
are assumed to be produced back to back and the non-photonic
$e^+-e^-$ background from uncorrelated pairs has been subtracted.}
\label{fig:correl}
\end{figure}

One could imagine that azimuthal correlations of non-photonic
$e^+-e^-$ pairs created in the decay of the heavy mesons whose heavy
quarks have been created together may carry information on the
energy loss mechanism. Many collisions with small momentum transfer
may better conserve the original back to back correlations than few
collisions with a large energy transfer. As displayed in
fig.\ref{fig:correl}, this is not the case. Model A and model E give
about the same azimuthal correlation. This means, on the other hand,
that correlations are a quite robust observable to test this
reaction scenario and to confront it with other ideas like the
AdS-CFT approach \cite{Horowitz:2008ig}.

\section{Conclusion and outlook}
In conclusion, we have found that it is possible to reduce the
uncertainties inherent in present day calculations of the energy
loss and of the $v_2(p_T)$ distribution of heavy quarks traversing a quark
gluon plasma by\\
a) determining the infrared regulator by the requirement
that it reproduces the energy loss calculated in the hard thermal
loop + semi-hard approach\\
b) using an effective infrared safe physical coupling constant which describes
other data like the gluon radiation in $e^+e^-$ annihilation and the
non strange decay of $\tau$ leptons.

Results of calculations in which these new features are employed
come close to the experimental data for $R_{AA}(p_T)$ as well as for
$v_2(p_T)$. The K factor required to reproduce the data is in
between 1.5 and 2. Up to now a simultaneous description of $R_{AA}$
and $v_2$ has not been possible even with large K-factors. That the
K-factor is above one may be due to radiative processes which are
not included here but it may also be due to the lack of a detailed
knowledge of the different physical processes involved. They include
the initial distribution of charm and bottom quarks, their
hadronization and the role of heavy baryons.

This observation has importance far beyond the physics of heavy
mesons. Because the same running coupling constant and the same
infrared regulator appear also in the cross section for light quarks
we expect a similar energy loss for light quarks. Pions show indeed
a very similar $R_{AA}(p_T)$ distribution but baryons do not. The
reason for this is unknown but if one follows the idea that they are
formed by coalescence their formation mechanism may be rather
different compared to that of heavy mesons. This conjecture is
supported by their large $v_2$ values. In addition, the large
collective radial flow counteracts to the individual energy loss. To
clarify the hadronization mechanism of light hadrons one has
probably to wait until jet like hadrons and those created by the
plasma hadronization can be separated, either by measuring
correlations or by extending the detection range in momentum space
in future LHC experiments.

The observed enhanced cross section may also be of importance
for the understanding of the fast equilibration
observed in entrance channel of ultrarelativistic heavy ion
collision where we do not have a heat bath like here but a momentum
distribution given by the structure functions. There the typical
momentum is, however, not far from that of the heat bath
particles.

\medskip

 {\bf Acknowledgments:} We thank A.~Peshier and
S.~Peign\'e for fruitful discussions and R. Vogt for communication
details of the approach of ref.\cite{Cacciari:2005rk}.

\section{Appendix}
\subsection{HTL+hard}
As the large transfer t will bring the parton to a final state
k' for which  $n_F(k')<<1$ we neglect the factor $1-n_F(k')$ for the
final state particle. We start from
 \bea -\left.\frac{dE_\mu}{dx}\right|^{v\rightarrow
1}_{|t|>|t^*|}&=& \int\frac{d^3k}{(2\pi)^3 2k} n_F(k) \int_{t_{\rm
min}}^{t^*} dt (-t) d_F\frac{d\sigma}{dt}\nonumber\\ &=&
\frac{d^3k}{(2\pi)^3 2k} n_F(k) \int_{t_{\rm min}}^{t^*} dt (-t)
\frac{1}{16\pi(s-M^2)^2}\,\frac{1}{d}\times 32 g^4
\left[\frac{(s-M^2)^2}{t^2}+\frac{s}{t}+\frac{1}{2}\right],
\nonumber\\
\eea where $M$ is the mass of the muon and $n_F$ is the Fermi-Dirac
distribution for a massless fermion. As
\begin{equation}
\frac{1}{(s-M^2)^2} \int_{t_{\rm min}}^{t^*} dt (-t)
\left[\frac{(s-M^2)^2}{t^2}+\frac{s}{t}+\frac{1}{2}\right] \approx
\ln\frac{|t_{\rm min}|}{|t^*|}-\frac{3}{4} \approx
\ln\frac{s}{|t^*|}-\frac{3}{4}
\end{equation}
for $s\gg M^2 \gg |t^*|$, we have
\begin{equation}
-\left.\frac{dE_\mu}{dx}\right|^{v\rightarrow 1}_{|t|>|t^*|}\approx
\frac{g^4}{16 \pi^4} \int \frac{k}{e^{k/T}+1}\,
\left(\ln\frac{s}{|t^*|}-\frac{3}{4}\right)\,dk\,d\Omega\,,
\label{hardpartPP}
\end{equation}
where $s=M^2+2 E k(1-\cos\theta(\vec{p},\vec{k}))$ and where the
integral is performed in principle over a domain such that $|t_{\rm
min}|\approx s \ge |t^*|$. For $E\gg M\gg |t^*|^{\frac{1}{2}}$, one
can nevertheless argue on a physical basis that there is enough
``hardness'' in almost every collision in order to fulfill this
condition and the domain in which this is not the case becomes
negligible. We will therefore integrate over the whole k space as
the integral converges. Introducing
$u=1-\cos\theta(\vec{p},\vec{k})\in[0,2]$, the angular integral
leads to
\begin{eqnarray}
\int d\Omega &\rightarrow& 2\pi \int_0^2 \left(\ln\frac{M^2+2Ek
u}{|t^*|}-\frac{3}{4}\right)\,du
\nonumber\\
&=&4\pi\left(\ln\frac{4 E k+M^2}{|t^*|}+\frac{M^2}{4Ek}\, \ln\frac{4
E k+M^2}{M^2}-1-\frac{3}{4}\right)\,. \label{intangul}
\end{eqnarray}
Substituting the variable $k$ by $x=k/T$, we obtain the expression
\begin{equation}
-\left.\frac{dE_\mu}{dx}\right|^{v\rightarrow 1}_{|t|>|t^*|}\approx
\frac{g^4 T^2}{4 \pi^3} \int \frac{x}{e^x+1}\, \left[\ln\frac{4ET
x+M^2}{|t^*|}-\frac{7}{4}+\frac{M^2}{4ET x}\, \ln\left(1+\frac{4 ET
x}{M^2}\right)\right]\,dx\,.
\end{equation}
Because $E$ is assumed to be $\gg M^2/T$ and because the integral is
dominated by intermediate values of $x$ ($x\approx 1$), one can
neglect the last term in the integrand and take $M=0$ in the first
term and one arrives at \bea
-\left.\frac{dE_\mu}{dx}\right|^{v\rightarrow
1}_{|t|>|t^*|}&\approx& \frac{g^4 T^2}{4 \pi^3} \int
\frac{x}{e^x+1}\, \left(\ln\frac{4ET
x}{|t^*|}-\frac{7}{4}\right)\,dx\,\nonumber\\&\approx& \frac{g^4
T^2}{48 \pi}\left[
\ln\frac{8ET}{|t^*|}-\gamma-\frac{3}{4}-\frac{\zeta'(2)}{\zeta(2)}\right]\,
, \eea  eq. 7 of ref. \cite{Braaten:1991jj}.
\subsection{effective IR regulator}
The t-integration of eq. \ref{eq:born} yields
\begin{eqnarray}
{\cal I}&=& \frac{1}{(s-M^2)^2} \int_{t_{\rm min}}^{0} dt (-t)
d_F\frac{d\sigma_F}{dt}
\nonumber\\
&=& \frac{1}{(s-M^2)^2}\int_0^{|t_{\rm min}|}\left(
-\mu^2*\frac{(s-M^2)^2}{(|t|+\mu^2)^2}+ \frac{(s-M^2)^2+\mu^2
s}{|t|+\mu^2}-s+\frac{|t|}{2} \right)d|t|
\nonumber\\
&=& \left(1+\frac{\mu^2 s}{(s-M^2)^2}\right)
\ln\left(1+\frac{(s-M^2)^2}{\mu^2 s}\right)-1+\frac{(s-M^2)^2}{4
s^2} -\underbrace{\frac{|t_{\rm min}|}{|t_{\rm
min}|+\mu^2}}_{\approx 1}\,,
\end{eqnarray}
and we obtain for the energy loss
\begin{equation}
-\left.\frac{dE_\mu}{dx}\right|^{v\rightarrow 1}_{\rm eff}\approx
\frac{g^4 T^2}{8 \pi^3}\int_0^{+\infty}\int_0^2
\frac{x}{e^{x}+1}\,{\cal I}(s)\,dk\,du\,, \label{dedxmodel}
\end{equation}
where $s=M^2+2 ET x u$. We first notice that $\frac{\ln(1+a)}{a}$ in
${\cal I}$ (with $a=\frac{(s-M^2)^2}{\mu^2 s}$) is maximal and
bounded at $a=0$ ($s=M^2$) and then decreases like $\mu^2/s\propto
\mu^2/ET$ for larger values of $s$. It then brings a contribution
$\propto \mu^2/ET$ that is subdominant at large energies. In this
regime, $s$ is $\gg M^2$ for most of the $(u,x)$ integration domain,
so that the 3rd term of ${\cal I}$ can be replaced by its asymptotic
$1/4$ value and $|t_{\rm min}|$ in the logarithm can be replaced by
$s$. Therefore,
\begin{equation}
{\cal I} \approx\ln\frac{s+\mu^2}{\mu^2}-\frac{3}{4}-1
\approx\ln\frac{s+\mu^2}{e \mu^2}-\frac{3}{4}\,,
\end{equation}
and one realizes that $-\left.\frac{dE_\mu}{dx}\right|^{v\rightarrow
1}_{\rm eff}$ is nothing but the hard contribution eq.
\ref{hardpartPP} with $|t^*|\rightarrow e \mu^2$ and $M^2\rightarrow
M^2+\mu^2\approx M^2$. We thus can read off the result directly from
eq. \ref{hardPP2}:
\begin{equation}
-\left.\frac{dE_\mu}{dx}\right|^{v\rightarrow 1}_{\rm eff}\approx
\frac{g^4 T^2}{48 \pi}\left[ \ln\frac{8ET}{e
\mu^2}-\gamma-\frac{3}{4}-\frac{\zeta'(2)}{\zeta(2)}\right]\,.
\end{equation}
and obtain eq. \ref{dedemod2}.

\end{document}